\documentclass[sigconf, screen]{acmart}
\AtBeginDocument{%
  \providecommand\BibTeX{{%
    \normalfont B\kern-0.5em{\scshape i\kern-0.25em b}\kern-0.8em\TeX}}}

\setcopyright{acmcopyright}
\copyrightyear{2022}
\acmYear{2022}
\acmDOI{10.1145/1122445.1122456}

\acmConference[Internetware'22]{Internetware'22: the 13th Asia-Pacific Symposium on Internetware}{June 11--12, 2022}{Hohhot, China}
\acmPrice{15.00}
\acmISBN{978-1-4503-XXXX-X/18/06}
\usepackage{enumitem}
\usepackage{multirow}
\usepackage[most]{tcolorbox}
\usepackage[ruled, linesnumbered]{algorithm2e}
\SetNlSty{bfseries}{\color{black}}{}
\usepackage{listings}
\usepackage{subcaption}
\usepackage{graphicx}
\usepackage{tikz}
\newcommand\encircle[1]{%
  \tikz[baseline=(X.base)] 
    \node (X) [draw, shape=circle, inner sep=0, scale=0.75] {\strut #1};}

\newcommand{\finding}[2]{
\begin{center}
\fcolorbox{black}{gray!10}{\parbox{.97\linewidth}{
\textbf{\textit{Answer to RQ{#1}:}}
{#2}
}}
\end{center}
}

\newcommand{\nBenchmark}{5}
\newcommand{\nFuzzCase}{5}
\newcommand{\nFuzzPair}{15}
\newcommand{\duration}{3h}
\newcommand{\nRepeat}{10}

\newcommand{\zest}{Zest}
\newcommand{\pmuzest}{P-$\mu$Zest}
\newcommand{\nmuzest}{N-$\mu$Zest}

\newcommand{\mfactor}{20}
\newcommand{\nSelect}{10}
\newcommand{\nThread}{11}

\newcommand{\fuzzPair}{$\{case, tech\}$}

\begin{document}

% \title{Mutation score guided fuzzing}
% \title{$\mu$Fuzz: Enhancing Coverage Guided Fuzzing with Mutation Analysis}
% \title{Enhancing Coverage Guided Fuzzing with Mutation Testing}
% \title{Enhancing Coverage Guided Fuzzing with Mutation Testing: An Empirical Study}
% \title{$\mu$Fuzz: Enhancing Fuzzing with Mutation Analysis}
% \title{Coverage Guided Fuzzing with Mutation Testing: An Empirical Study}
% \title{Opportunities for Mutation Testing in Coverage Guided Fuzzing}
\title{Investigating Coverage Guided Fuzzing with Mutation Testing}

\author{Ruixiang Qian}
\authornote{Both authors contributed equallny to this research.}
\email{qrx@smail.nju.edu.cn}
\affiliation{%
  \institution{State Key Laboratory for Novel Software Technology}
  \city{Nanjing University}
  \country{China}
}

\author{Quanjun Zhang}
\authornotemark[1]
\email{quanjun.zhang@smail.nju.edu.cn}
\affiliation{%
  \institution{State Key Laboratory for Novel Software Technology}
  \city{Nanjing University}
  \country{China}
}

\author{Chunrong Fang}
\authornote{Corresponding author.}
\email{fangchunrong@nju.edu.cn}
\affiliation{%
  \institution{State Key Laboratory for Novel Software Technology}
  \city{Nanjing University}
  \country{China}
}

\author{Lihua Guo}
\email{garyglh@163.com}
\affiliation{%
  \institution{State Key Laboratory for Novel Software Technology}
  \city{Nanjing University}
  \country{China}
}


\begin{abstract}
\textit{Coverage guided fuzzing} (CGF) is an effective testing technique which has detected hundreds of thousands of bugs from various software applications.
It focuses on maximizing code coverage to reveal more bugs during \textit{fuzzing}. 
However, a higher coverage does not necessarily imply a better fault detection capability.
Triggering a bug involves not only exercising the specific program path but also reaching interesting program states in that path. 

In this paper, we use mutation testing to improve CGF in detecting bugs.
We use mutation scores as additional feedback to guide fuzzing towards detecting bugs rather than just covering code.
To evaluate our approach, we conduct a well-designed experiment on \nBenchmark{} benchmarks.
We choose the state-of-the-art fuzzing technique Zest as baseline and construct two modified techniques on it using our approach.
The experimental results show that our approach can improve CGF in both code coverage and bug detection.

\end{abstract}




\maketitle

% （1）现有基于覆盖模糊技术，大多使用分支覆盖进行引导；（2）但是现有研究表明，覆盖与缺陷检测之间存在一些差距，覆盖缺陷语句不一定可以表明检测到缺陷，（3）而在一些其他的领域，变异作为一种新的准则可以更好的度量用例的缺陷检测能力，因此我们通过对源码进行变异分析，以此作为一种覆盖的增强方案，作为度量种子的缺陷检测能力
\section{Introduction}
% Introduce fuzzing (Brief).
Fuzzing is one of the most popular techniques to test software correctness and reliability \cite{DBLP:conf/icse/GaneshLR09, fuzzSurvey}.
At high level, fuzzing is a process that repeatedly runs the program under test (PUT) with a mess of generated inputs, some of which maybe syntactically or semantically invalid. 
It relies on a component named \textit{fuzzer} to generate test inputs and execute the PUT.
Generally, a fuzzer generates inputs from given seeds and exercises the PUT continuously with the aim of exposing errors of PUTs in a period of time. 
To date, fuzzing is almost the most widely-adopted technique due to its conceptual simplicity, low barrier for deployment, and efficacy in discovering real-world bugs \cite{fuzzSurvey}.

% Introduce CGF (Detailed, more). Emphasize strengths of CGF: core: CGF focuses on maximizing code coverage. CGF does detect crashes and bugs. Some SOTA, show their achievements.
However, fuzzing can be rather ineffective at exploring different program paths.
This is because a fuzzer, or more specifically a \textit{black-box fuzzer}, exercises a PUT in a totally blind manner. 
White-box fuzzers adopt systematic effort\cite{SAGE, DBLP:conf/icse/GaneshLR09} to aid fuzzing in searching diverse program paths.
However, white-box fuzzers can slow down the execution of the PUT due to the heavy program analysis it performs \cite{fuzzSurvey}.
To make balance, \textit{coverage guided fuzzing} (CGF) \cite{aflfast, fuzzSurvey} uses lightweight instrumentation to gain coverage information from executions and guide fuzzing towards rarely executed paths with attained coverage.
Owing to its strengths, many CGF techniques have emerged in recent years\cite{afl, aflfast, zest, jqf, perffuzz, muzz}. 
For example, Zest \cite{zest} focuses on programs with syntax checks and has revealed semantic bugs in 5 widely used third-party applications; 
AFL \cite{afl} has been reported to detect tens of thousands of vulnerabilities in hundreds of real-world software projects \cite{muzz}.

% Outer challenges: Drawbacks of CGF => strengths of our method
% \qrx{Not direct enough} \qrx{coverage and fault detection.} 
% \qrx{Target assumption not bugs...}
CGF endeavours to reveal bugs within the PUT via maximizing code coverage \cite{mutestsurvey}.
However, a higher coverage does not necessarily imply a better bug detection capability.
Many researches have revealed that the correlation between code coverage and bug detection capability is weak \cite{DBLP:journals/tosem/GligoricGZSAM15, DBLP:conf/qrs/Hemmati15, DBLP:journals/jss/MaCZX20}.
Concentrating only on code coverage provides inadequate feedback for fuzzing, which may potentially lower its effectiveness in detecting bugs.

In this paper, we incorporate CGF with mutation testing to address the aforementioned challenges.
\textit{Mutation testing} is a fault-based testing technique which realises the idea of using \textit{mutants} (buggy versions mutated from the PUT) to support testing activities \cite{mutestsurvey}.
We exploit mutation testing to identify bug-revealing inputs generated by fuzzers, and use mutation scores as additional feedback to guide fuzzing towards detecting bugs. 
Our approach comprises two stages: (1) in initialization stage, we create mutants and make preparations for fuzzing, and (2) in fuzzing stage, we generate inputs and execute them against PUT and mutants.
We check fault detection capability for each generated test input and preserve the bug-revealing ones, i.e., the inputs that can kill any mutants.

We conduct a well-designed experiment with \nFuzzPair{} fuzz campaigns involving \nBenchmark{} benchmarks to evaluate the proposed approach. 
We choose the state-of-the-art CGF technique Zest \cite{zest} as baseline and construct two modified techniques on it with our approach in both 
\textit{negative} (generating fewer children from inputs kill mutants which have been killed in previous) and 
\textit{positive} (generating more children from inputs which are capable of killing any mutants)
manners.
The experimental results show that mutation testing can help CGF
(1) to maximize code coverage faster in 3 out of 5 benchmarks with a negative manner, and 
(2) to reveal bugs faster in all \nBenchmark{} benchmarks.
Besides, the techniques modified with our approach detects 10 more bugs in one of the benchmarks.
We summarize the main contributions of this paper are as follows:
\begin{itemize}[leftmargin=*, topsep=3pt]
    \item \textbf{Novel Approach.}
    We propose an approach which uses mutation scores as feedback to guide fuzzing towards detecting bugs.
    To our best knowledge, this is the \textbf{\textit{first}} work which incorporates CGF with mutation testing.
    % We create mutants before fuzzing, and select part of the created mutants to compute mutation scores during fuzzing.
    % We use mutation scores as feedback to guide fuzzing towards detecting bugs.
    
    \item \textbf{Practical Framework.}
    % We implement the proposed approach as a framework composed of three major components: 
    We implement the proposed approach as a framework which comprises a \textit{mutation engine}, a \textit{testing engine} and a \textit{fuzzing engine}.
    The components of our framework are scalable and can be generalized to other CGF techniques.
    
    \item \textbf{Extensive Study.}
    We conduct an experiment with \nFuzzPair{} fuzz campaigns involving \nBenchmark{} benchmarks.
    The experimental results demonstrate that our approach outperforms CGF in both code coverage and bug detection.
    
\end{itemize}

\section{Background}
\subsection{Coverage Guided Fuzzing}
\label{subsec:cgf}

\begin{algorithm} [ht]
    \caption{Coverage Guided Fuzzing}
    \label{alg:cgf}
    \KwIn {program $p$, initial inputs $I$}
    \KwOut {seed inputs queue $Q$, failing inputs set $F$}
    \label{fuzz_start}
    $Q \leftarrow I$ \; 
    $F \leftarrow \emptyset$ \;
    $Cov \leftarrow \emptyset$ \;
    \Repeat{exceeding given resources} { \label{fuzz_loop_start}
        \ForEach{input $i$ in Q} {
            $n \leftarrow$ \texttt{mutationChance($i, Q$)} \;
            \For{$0 < i < n$}{
                $i_c \leftarrow$ \texttt{mutate($i, Q$)} \; \label{mutate_input}
                $cov, res \leftarrow$ \texttt{execute($p, i_c$)}\;
                \uIf{\texttt{isCrash($res$)}}{
                    $F \leftarrow F \cup \{i_c\}$ \;
                }
                \ElseIf{\texttt{existNewCov($cov$)}}{
                    $Q \leftarrow Q \cup \{i_c\}$ \; 
                    $Cov \leftarrow Cov\cup \{cov\}$ \;
                }
            }
        }
    } \label{fuzz_loop_end}
    \Return{$Q, F$} \; 
\end{algorithm}

\noindent 
Algorithm \ref{alg:cgf} presents the typical process of CGF.
It takes a instrumented program $p$ and a set of initial seeds $I$ as inputs, and return saved seeds $Q$ as well as seeds lead to crashes $F$ as outputs.
At first, a coverage guided fuzzer add all inputs given in $I$ to $Q$ and initialize $F$ and global coverage $Cov$ as empty.
Next, it comes into the main fuzzing loop (line \ref{fuzz_loop_start}  \textasciitilde{} line \ref{fuzz_loop_end}).
For each input $i$ in $Q$, the fuzzer computes the number of mutations $n$ it going to perform on current input (line \ref{mutate_input}).
The computation heuristic is abstracted as a function \texttt{mutationChance}, whose process is shown as Algorithm \ref{alg:mc}.
Next, the fuzzer mutates current input $i$ for $n$ times to get children inputs $i_c$.
For each $i_c$, the fuzzer executes $p$ with it once to check the attained coverage $cov$ and execution result $res$.
If the $res$ is failure, then $i_c$ is a failing input and will be added into $F$.
This suggests that $i_c$ has crashed the execution and triggered some vulnerabilities of $p$;
If the $res$ is success, the fuzzer further checks whether $cov$ contains some paths that didn't covered before.
If new coverage attained, then $i_c$ will be added into $Q$ for subsequent fuzzing, and the total coverage $Cov$ will be updated.

\begin{algorithm} [ht]
    \caption{Computation of Mutation Chance}
    \label{alg:mc}
    \KwIn {seed input $i$, seed inputs queue $Q$}
    \KwOut {number of mutations $n$}
    $n \leftarrow$ \textit{BASE}\;
    \If{\texttt{canProduceNewCov($i, Q$)}}{ \label{favor_s}
        $n \leftarrow n$ $\times$ \textit{FACTOR}\;
    }  \label{favor_e}
    \Return{n}
\end{algorithm}

In Algorithm \ref{alg:mc}, both \textit{BASE} and \textit{FACTOR} are positive constants.
To begin with, the number of mutations for each input $i$ is set to the unified baseline constant $BASE$.
Next, to show the favor for inputs that can supply new coverage, another constant $FACTOR$ is used to scale up the numbers of mutations for these inputs through multiplication (line \ref{favor_s} \textasciitilde{} \ref{favor_e}).
The intuition behind these manipulations is to keep the quality of generated inputs with the help of "Matthew Effect" \cite{muzz}.
However, traditional CGF use same heuristic to compute $n$ for every inputs.
It does not distinguish inputs that are capable of detecting bugs from normal ones, which consequently hinder fuzzing from finding more bug-revealing inputs.
We will elaborate the challenges and show our solution at Section \ref{subsec:motivation}.

\subsection{Mutation Testing}
\label{subsec:mutation_testing}

\noindent
Mutation testing is a fault-based testing technique \cite{mutestsurvey, pitest}.
It uses artificial bugs, called mutants, to evaluate the adequacy of testing activities.
% The general mutation testing process is shown at Figure \ref{fig:mutation_testing}.
Given a PUT and a set of test inputs, mutation testing firstly generates a set of mutants with a mutation engine, and then executes these test inputs against these mutants to compute mutation score.

A mutation engine generate mutants in three steps: 
Firstly, it selects a set of mutators (syntactic rules which encodes the transformation of the syntax of program) to create mutants.
Secondly, it creates a group of mutants according to used mutators.
Thirdly, it optimizes mutants through removing redundant mutants.
Note that mutant creation may result in mutants that are equivalent to or subsumed by other mutants \cite{mutestsurvey}.
Besides, some mutants may semantically equivalent to the PUT, even though they are syntactically different.
These mutants are detrimental to the result of mutation result such that should be removed before execution \cite{mutestsurvey}. 

% \begin{figure}[htb]
%     \centering
%     \includegraphics[width=0.47\textwidth]{pic/mutation_testing_workflow.pdf}
%     \caption{Workflow of mutation testing}
%     \label{fig:mutation_testing}
% \end{figure}

The generated mutants are then executed to evaluate the adequacy of test inputs.
The adequacy of test inputs is measured by mutation score, which can be computed as follows:
\begin{equation*}
\label{eq:mutation_score}
    score = \frac{mut_k}{mut_s + mut_k} \times 100\%
\end{equation*}
Specifically, mutation score is the ratio of killed mutants ($mut_k$) to all mutants (the sum of killed and survived mutants).
If the execution of a mutant with the given test inputs \textit{fails}, it means the defect represented by this mutant is detected such that the mutant is \textit{killed};
Otherwise the mutant is \textit{survived}, implying that the given inputs are incapable of detecting such a defect.
Generally, mutation score can be used to reflect the capability of given inputs in detecting bugs.
Suppose only one input was sent to mutation testing, then a non-zero mutation score implies that the input is capable of detecting bugs.

\subsection{Motivation}
\label{subsec:motivation}

\begin{lstlisting}[
    language={java},
    label={list:foo},
    caption={A simple method that is supposed to return the larger value of \texttt{x} and \texttt{y}. A bug lies on line \ref{line:bug} where the \texttt{y} is larger one and should be returned.}, 
    captionpos={b},
    % linewidth=10cm,
    escapechar={|} % for labeling line
]
int foo(int x, int y) {
    if (x > y)
        return x; 
    else
        return x; // Should return y. | \label{line:bug} |
}
\end{lstlisting}

\noindent
\textbf{Motivating example.}
Consider the code snippet presented at List \ref{list:foo}. 
The method \texttt{foo} takes two integers \texttt{x} and \texttt{y} as inputs and is supposed to return the larger value among them as output.
However, there is a bug lies on line \ref{line:bug} in that \texttt{else} clause.
It is \texttt{y} rather than \texttt{x} that should be returned as \texttt{y} is no less than \texttt{x}.

This code snippet illustrates the situation that a bug is reached but may not be triggered.
Consider we have two test inputs $i_1 = \langle 1, 1 \rangle$ and $i_2 = \langle 1, 2 \rangle$.
Both of $i_1$ and $i_2$ cover line \ref{line:bug} but only $i_2$ can trigger the bug by making it observable at output.

\noindent
\textbf{Drawbacks of traditional CGF.}
Suppose test input $i_1$ and $i_2$ are generated in sequence during a fuzz campaign.
According to Algorithm \ref{alg:cgf}, traditional CGF will not preserve $i_2$ as it supplies same coverage as $i_1$. 
As a result, the bug-revealing input $i_2$ is discarded, and the fault lies at line \ref{line:bug} may hidden.

\begin{lstlisting}[
    language={java},
    label={list:foo_mutant},
    caption={A mutant of code snippet illustrated at List \ref{list:foo} where a ``ReturnZero'' mutator is conducted at line \ref{line:bug}}, 
    captionpos={b},
    escapechar={|} % for labeling line
]
int foo(int x, int y) {
    if (x > y)
        return x;  
    else
        |\textcolor{blue}{\texttt{\textbf{return }0;}}| // Being mutated. | \label{line:mut} |
}
\end{lstlisting}

\noindent
\textbf{Our solution.}
We leverage mutation testing to address the drawbacks of traditional CGF.
We firstly create a mutant for PUT, which is shown in List \ref{list:foo_mutant}.
Specifically, this mutant is created by conduct a ``ReturnZero'' mutator at line \ref{line:mut} of the method \texttt{foo}. 
After that we perform mutation testing with the created mutant. 
In addition to the original PUT (List \ref{list:foo}), we execute the created mutant with the same inputs ($i_1$ and $i_2$) generated during fuzzing.
We check the consistency between the outputs of original PUT and the created mutant to compute mutation score.
If outputs are consistent, then the mutant is survived and the mutation score is ``0'';
Otherwise the mutant is killed and the mutation score is ``1''.
We prefer bug-revealing inputs such that preserve the inputs which are capable of killing the mutant (getting a mutation score of ``1'').
With our approach, the bug-revealing input $i_2$ will not be discarded as before, and the bug lies at Lsine \ref{line:bug} can be potentially detected. 

\section{Approach}
\label{section:approach}
% Additional parts to general CGF
\newcommand{\algomod}[1]{\textcolor{blue}{#1}}
% Short name for fault detetion aware CGF
\newcommand{\algoname}{$\mu$CGF}
\newcommand{\mutengine}{\encircle{\textbf{M}}}
\newcommand{\testengine}{\encircle{\textbf{T}}}
\newcommand{\fuzzengine}{\encircle{\textbf{F}}}
\newcommand{\valengine}{\encircle{\textbf{V}}}
\newcommand{\selthreshold}{10}
% Define factors
\newcommand{\killfac}{\textit{KILL\_FACTOR}}
\newcommand{\killnewfac}{\textit{KILL\_NEW\_FACTOR}}

\subsection{Overview}
\begin{figure*}[htb]
    \centering
    \includegraphics[width=0.98\textwidth]{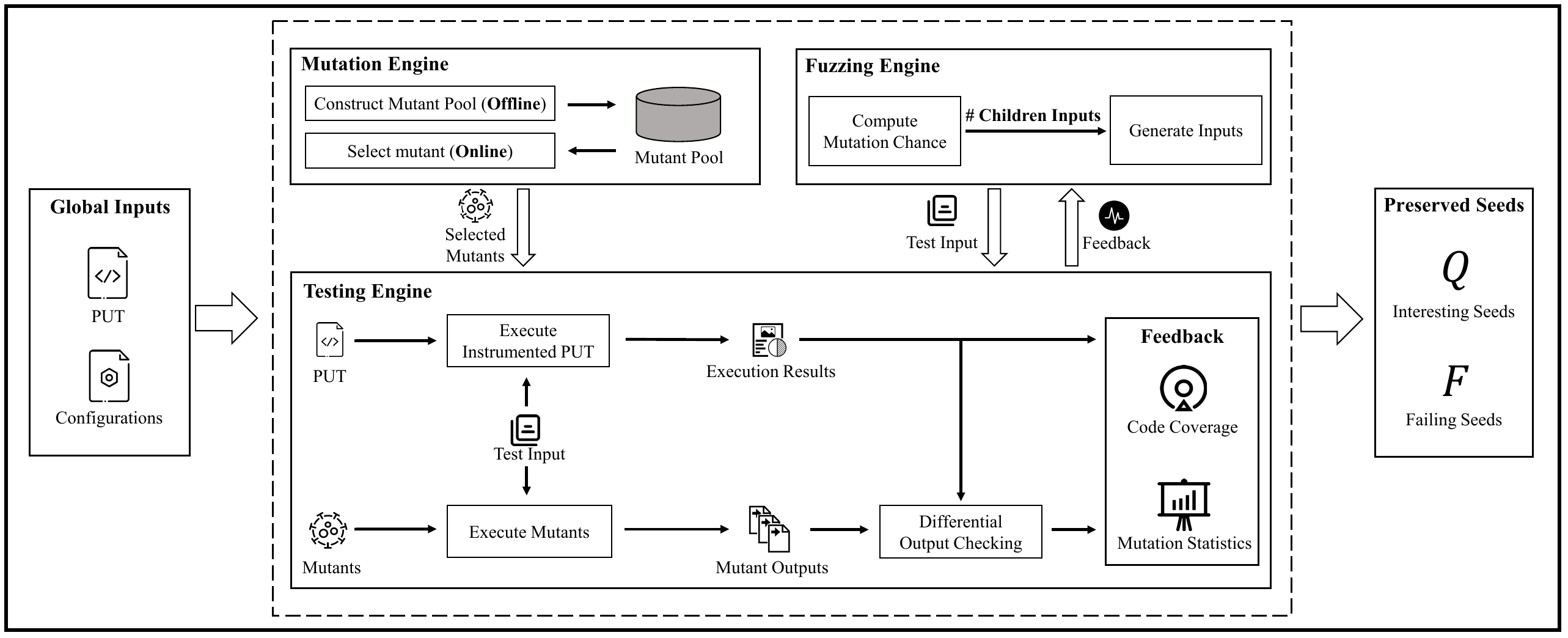}
    \caption{Overview of our approach.}
    \label{fig:overview}
\end{figure*}

\begin{algorithm} [htp]
    \caption{Fault Detection Aware CGF}
    \label{alg:modified_cgf}
    \KwIn {program $p$, initial inputs $I$, \algomod{mutation configuration $c$}}
    \KwOut {seed inputs queue $Q$, failing inputs set $F$}
    $Q \leftarrow I$ \; 
        $F \leftarrow \emptyset$ \;
        $Cov \leftarrow \emptyset$ \;
        \algomod{$M \leftarrow $ \texttt{buildMutantPool($p, c$)}} \;
        \Repeat{exceeding given resources} {
            \ForEach{input $i$ in Q} {
                $n \leftarrow$ \texttt{mutationChance($i, Q$)} \;
                \For{$0 < i < n$}{
                    $i_c \leftarrow$ \texttt{mutate($i, Q$)} \; 
                    \algomod{$M' \leftarrow$ \texttt{selectMutants($M, c$)}} \; \label{line:mutant_selection}
                    $cov, res, \algomod{stat} \leftarrow$ \texttt{execute($p, \algomod{M'}, i_c$)}\;
                    \algomod{
                    \If{\texttt{canKillMutants($stat$)}}{ \label{line:check_kill}
                        \texttt{markAsCapable($i_c$)} \;\label{line:mark_capable}
                        $Q \leftarrow Q \cup \{i_c\}$ \;
                        \If{\texttt{canKillNewMutants($stat, M$)}}{ \label{line:check_kill_new}
                            \texttt{markKillNew($i_c, stat$)}\; 
                            \texttt{update($M, stat$)}\; \label{line:update_M}
                        }
                    }
                    }
                    \uIf{\texttt{isCrash($res$)}}{
                        $F \leftarrow F \cup \{i_c\}$ \;
                    }
                    \ElseIf{\texttt{existNewCov($cov$)}}{
                        $Q \leftarrow Q \cup \{i_c\}$ \; 
                        $Cov \leftarrow Cov\cup \{cov\}$ \;
                    }
                }
            }
        } 
        \Return{$Q, F$} \;
\end{algorithm}
\begin{algorithm} [htb]
    \caption{Fault Detection Aware Mutation Chance}
    \label{alg:modified_mc}
    \KwIn {seed input $i$, seed inputs queue $Q$}
    \KwOut {number of mutations $n$}
    $n \leftarrow$ \textit{BASE}\;
    \If{\texttt{canProduceNewCov($i, Q$)}}{ 
        $n \leftarrow n$ $\times$ \textit{FACTOR}\;
    }
    \algomod{
    \uIf{\texttt{isBugRevealing($i$)}}{
        $n \leftarrow n$ $\times$ \killfac{}\;
    }
    \Else{
        $n \leftarrow n$ $\div$ \killfac{}\;
    }
    }
    \Return{n}
\end{algorithm}

Figure \ref{fig:overview} depicts the overview of our approach.
Our approach takes a PUT and system configurations as inputs and outputs preserved seeds (i.e., test inputs).
System configurations not only contain settings for mutation testing but also settings for fuzzing such as allocated resources and initial seeds. 
We elaborate modified CGF algorithms (Algorithm \ref{alg:modified_cgf} and Algorithm \ref{alg:modified_mc}) in Section \ref{subsec:aware_cgf}.
Our approach comprises three main components: 
(1) a \textit{mutation engine} \mutengine{} for creating and selecting mutants (Section \ref{subsec:mutant_creation_selection}), 
(2) a \textit{testing engine} \testengine{} for executing programs and computing feedback (Section \ref{subsec:test_execution}), and
(3) a \textit{fuzzing engine} \fuzzengine{} for computing mutation chances and generating inputs (Section \ref{subsec:input_gen}). 

\subsection{Fault Detection Aware CGF}
\label{subsec:aware_cgf}

We modified the general CGF workflow to enable (1) the identification of bug-revealing inputs and (2) the preservation of fault detection capability.
Algorithm \ref{alg:modified_cgf} shows the \textit{fault detetion aware} CGF (\algoname{} in short) modified from Algorithm \ref{alg:cgf}.
The additional parts are highlighted in blue.
In spite of the PUT $p$ and initial inputs $I$, \algoname{} requires a configuration $c$ for mutation testing.
It comprises a set of mutators for mutant creation before fuzzing and a strategy for mutation selection (line \ref{line:mutant_selection}) during fuzzing.
Before coming into main fuzz loop, we additionally create a \textit{mutant pool} $M$ for $p$ according to $c$ as preparation.
During fuzzing, we perform mutation testing with each generated input $i_c$ by executing it against selected mutants $M'$ to get statistics $stat$.
Note that we only select partial mutants to ensure the efficiency of fuzzing.
By $stat$ we record detailed information to indicate whether a specific mutant was killed or survived in the last execution.  
We identify whether a generated $i_c$ is bug-revealing via checking $stat$ at line \ref{line:check_kill}.
Specifically, an $i_c$ is capable of detecting bugs if it has gotten a non-zero mutation score in the last execution.
For the $i_c$ capable of detecting bugs, we mark it as ``capable'' and preserve it into seed queue $Q$.
After that, we further check whether $i_c$ can kill new mutants (which has not been killed by any previous inputs yet) at line \ref{line:check_kill_new}.
For $i_c$ killed new mutants, we mark it in $stat$ and update the statuses for each mutants in $M$ at line \ref{line:update_M}.

With Algorithm \ref{alg:modified_cgf} the bug-revealing inputs are marked and now identifiable.
On this basis, we further propose Algorithm \ref{alg:modified_mc} to preserve fault detection capability of these bug-revealing inputs.
We realize this through rewarding bug-revealing inputs by amplifying their chances of being mutated. 
The inputs which are not bug-revealing will be penalized in an opposite way.
Note that the ``mutation'' here indicates the process of generating children inputs from parent seeds. 
We set a positive constant \killfac{} to adjust the computation of mutation chances.
At first the number of mutations $n$ is assigned by a fixed constant $BASE$.
If an inputs it bug-revealing, we multiply $n$ with \killfac{} to amplify its influences in the subsequent input generation.
Otherwise, we avoid generating children inputs from it by dividing its $n$ with \killfac{}.
Note that it is possible that an actually bug-revealing input is marked as ``incapable'' as we only select a part of mutants from $M$ during fuzzing due to efficiency concern.
The mutants the input can kill may not be selected thus the input will be not be treated as bug-revealing. 
% We believe that it is acceptable to misjudge only part of the bug-revealing inputs.
The trade of missing some bug-revealing inputs for efficiency is acceptable as the main goal of Algorithm \ref{alg:modified_cgf} is to preserve fault detection capability. 
With Matthew effect \cite{muzz}, only a small faction of bug-revealing inputs preserved can improve the possibility of generating bug-revealing children in numerous cycles of fuzzing.

\subsection{Mutant Creation and Selection}
\label{subsec:mutant_creation_selection}

\mutengine{} provides two modes: an \textit{offline mode} for mutant creation and a \textit{online mode} for on-the-fly mutation selection. 
The \textit{offline mode} 
is used for preparing fault detection aware fuzzing.
Mutants are created according to mutators specified in mutation configuration $c$ (Algorithm \ref{alg:modified_cgf}).
We preserve a mutated version of the PUT together with corresponding mutation information (such as mutated location and used mutator) for each mutant, and construct a mutant pool for subsequent usage.
The \textit{offline mode} is performed only once to avoid overheads caused by repeated creations.
On the contrary, the \textit{online mode} continuously works during test executions.
\mutengine{} periodically selects a subset of mutants from the pool according to a selection strategy specified in $c$. 
The selected mutants are sent to \testengine{} to for mutation testing.
Generally, by selection strategy, we only pick a small set (typically \selthreshold{}) of mutants to ensure the efficiency of fuzzing as the number of total mutants can be very large \cite{mutestsurvey, pitest}, especially when there are too many locations for mutation.
% The trade-off?

\subsection{Test Execution}
\label{subsec:test_execution}
\testengine{} is the core of our approach.
It extends the test execution component of traditional coverage guided fuzzers. 
\testengine{} receives test inputs from \fuzzengine{} and sends feedback information, i.e. coverage and mutation statistics, back to \fuzzengine{} to aid input generation.
Besides executing the instrumented PUT with the received test inputs, \testengine{} also needs to execute the same inputs against the mutants selected by \mutengine{}.

% Besides executing instrumented PUT to collect coverage and trigger vulnerabilities, \testengine{} executes the same input against the mutants selected by \mutengine{} to get statistics about fault detection.
% \testengine{} receives test inputs from \fuzzengine{} and sends feedback information, i.e. coverage and fault detection statistics back to \fuzzengine{} to aid input generation.

\newcommand{\success}{\texttt{SUCCESS}}
\newcommand{\failure}{\texttt{FAILURE}}
\newcommand{\survived}{\texttt{SURVIVED}}
\newcommand{\killed}{\texttt{KILLED}}

\newcommand{\testResult}{test result}
\newcommand{\mutestResult}{mutation test result}

% tuple <>, used in math environment
\newcommand{\tuple}[1]{\langle #1 \rangle}
% set {}, %%used in math environment
\newcommand{\set}[1]{\{ #1 \}}

\newtheoremstyle{mydef}% name
{3pt}% Space above
{3pt}% Space below
{\upshape}% Body font
{}% Indent amount
{}% Theorem head font
{.}% Punctuation after theorem head
{.5em}% Space after theorem head
{}% Theorem head spec (can be left empty, meaning ‘normal’ )
\theoremstyle{mydef} % Param is the name of theorem
\newtheorem{mydef}{\textbf{\textsc{Definition}}}

% \subsubsection{Formal Definitions}
\subsubsection{Terms}
We formalize the terms used in this section to avoid ambiguous understanding.
% To simplify representation, we firstly make some definitions for terms used in this section.
\begin{mydef}[\textit{Program}]
    A program is an object of test. In this paper, a program can be a PUT $p$ or a mutant $mut$.  
\end{mydef}
\begin{mydef}[\textit{Program Behavior}]
    A program behavior represents a functionality of a PUT, which can be denoted as $b$.
    A PUT involves a set of behaviors $\mathbb{B} = \{b_1, b_2, ..., b_n\}$, where $n \geq 1$.
\end{mydef}
\begin{mydef}[\textit{Output}]
    An output $o$ is the result of exercising a program behavior $b$ belongs to a program $p$ with a given test input $i$, which can be denoted as $o = b[p, i]$. 
\end{mydef}
\begin{mydef}[\textit{Execution Result}] \label{def:exec_res}
    An execution result $exec$ is a triple $\langle s, \mathbb{O}, e \rangle$, where 
    $s$ is the status of execution which satisfies $s \in  \{\success, \failure\}$, 
    $\mathbb{O}$ is a set of outputs which can be denoted as $\mathbb{O} = \{o_1, o_2, ..., o_n\}, (n \leq 1)$, and
    $e$ is the crash occurred during execution.
    Note that one of $\mathbb{O}$ and $e$ is null as when $s$ is \success{}, the execution was successful such that no crash will be raised; whereas in turn, when $s$ is \failure{}, the execution was crashed and no outputs will be produced.
\end{mydef}
\begin{mydef}[\textit{Mutant Status}]
    A mutant status represents whether a artificial bug represented by a mutant $mut$ was detected, which can be denoted as $\epsilon[mut]$.
    A mutant status $\epsilon$ satisfies $\epsilon[mut] \in \{\texttt{KILLED}, \texttt{SURVIVED}\}$.
    We further discuss the judging of mutant statuses in Section \ref{subsubsec:mutation_killing_framework}.
\end{mydef}

\subsubsection{Running Mutants and PUT}
Mutation testing can be very time consuming \cite{mutestsurvey, pitest} as it has to run a single test input against every selected mutants once.
Note that we do not distinguish \textit{mutation analysis} and \textit{mutation testing} \cite{mutestsurvey} as there is only one test input to run at each execution. 
We have reduced the number of mutants to execute by strategically selecting a subset of mutants (Section \ref{subsec:mutant_creation_selection}).
To further improve execution efficiency, we concurrently execute the selected mutants together with the PUT in multi-threads. 
We collect execution results for both PUT and the selected mutants.
Herein, we call execution results for PUT and mutation testing as ``\testResult{}'' and "\mutestResult{}" respectively.
A test execution is \failure{} if it was crashed (e.g. throws an exception) during execution, or else it is \success{} (Definition \ref{def:exec_res}). 
We allow mutants to run slightly longer than the PUT does, and provide an \textit{extra execution time ratio} to control the extra time.
We make this design for two reasons:
(1) To get more confidential mutation statistics.
Test executions can accidentally fluctuate due to reasons like system scheduling.
Besides, since mutants are created using various mutators (each mutator is a kind of syntactic transformation), some mutants exactly need more time to execute than the others.
A longer time can enable more mutants to finish executions without timeout, which consequently reduces false negatives;
(2) To keep fuzzing efficient.
Syntactic transformations on the PUT may sometimes result in time-consuming artificial bugs such as endless loop in practice \cite{pitest}.
Setting an upper bound time for executing mutants can prevent executions from being blocked by such unexpected issues.
% For a PUT whose execution time is $t$, we use extra ratio $r(0 \leq r \leq 1)$ to control the upper bound time  $t'$ for executing mutants, which is computed as:
For a PUT whose execution time is $t$, we use extra ratio $r(0 \leq r \leq 1)$ to control the upper bound time  $t'$ for executing mutants, which is computed as $t' = t \times (1 + r)$.
% \begin{equation*}
%     t' = t \times (1 + r)
% \end{equation*}
Note that we consider the \mutestResult{}s of mutants which does not finished executing in $t'$ as \failure{} during fuzzing.

\subsubsection{Analyzing Execution Results}
\label{subsubsec:mutation_killing_framework}
Test execution results should be further analyzed to identify the nature of test inputs.
Through such analysis, 
We further analyze test execution results to identify whether a test input has supplied fresh coverage or killed mutants.
The the process of analysis varies according to execution statuses.
Specifically, if \testResult{} is \failure{}, the test input is failing input should be saved into failing seed queue $F$ (Algorithm \ref{alg:modified_cgf}).
Mutation test results will not be analyzed no outputs was produced;
Else, if \testResult{} is \success{}, we further analyze execution results to determine the statuses of mutants.
A mutant is killed if it is detected by certain test cases (Section \ref{subsec:mutation_testing}).
Killing mutants requires the buggy outputs of mutants can be observed explicitly.
However, as executed test inputs are generated dynamically during fuzzing, the real expected outputs are impossible to know in advance.
Therefore, we use output of the PUT as expected output.
We build a \textit{mutant checking mechanism} based on the thoughts of  \textit{differential testing}.
Differential testing \cite{98difftest} is a random testing technique which reveals potential bugs through comparing the outputs (with same inputs) of among comparable software systems.
% It addresses the cost of evaluating test results.
We treat a mutant as a comparable system to the PUT and identify the statuses of it by comparing its outputs $\mathbb{O'}$ to the outputs of the PUT $\mathbb{O}$.
We formalize the goal of the mutant checking as follows:
\begin{mydef}[\textit{Differential Outputs Checking}]
    Given a test input $i$, a set of program behaviors $B = \{b_1, b_2,... , b_m\}$, a PUT $p$ and a set of selected mutants $M' = \{mut_1, mut_2,..., mut_n\}$, the goal of \textit{mutant checking} is to determine status for each $mut \in M'$ via comparing its outputs $\mathbb{O'}$ to outputs $\mathbb{O}$ of the PUT $p$.
    The result of mutant checking are mutation statistics $stat$ described in Algorithm \ref{alg:modified_cgf}.
\end{mydef}

Specifically, if the execution of a mutant $mut$ has crashed, or some of its outputs are inconsistent to the PUT, then the mutant is detected and \textit{killed} by the executed test input $i$.
Otherwise, $mut$ is ignored such that \textit{survived}.
We define mutant surviving and killing as follows:
% Specifically, if the execution of a mutant $mut$ has crashed, or some of its outputs are inconsistent to the PUT, then the mutant is detected and \textit{killed} by the executed test input $i$.
% Otherwise, $mut$ is ignored such that \textit{survived}.
% We define mutant surviving and killing as follows:
\begin{mydef}[\textit{Mutant Surviving}]
    Given a PUT $p$ whose \testResult{} is $exec = \tuple{s, \mathbb{O}, e}$ and
    a mutant $mut$ whose \mutestResult{} is $exec' = \tuple{s', \mathbb{O'}, e'}$, 
    the $mut$ is \textit{survived} if the execution status is \success{} and \textbf{all} its outputs are consistent to the PUT. 
    We denote \textit{mutant surviving} as $s' = \success \wedge \mathbb{O} = \mathbb{O'} \Rightarrow \epsilon[mut] = \survived$.
\end{mydef}
\begin{mydef}[\textit{Mutant Killing}]
    Given a PUT $p$ whose \testResult{} is $exec = \tuple{s, \mathbb{O}, e}$ and
    a mutant $mut$ whose \mutestResult{} is $exec' = \tuple{s', \mathbb{O'}, e'}$,
    the $mut$ is \textit{killed} if the execution status is \failure{} or \textbf{existing} at least one output which is inconsistent to that of $p$.
    We denote \textit{mutant killing} as $s' = \failure{} \vee \exists j \rightarrow o_j \neq o'_j, o_j \in \mathbb{O}, o'_j \in \mathbb{O'} \Rightarrow \epsilon[mut] = \killed$.
\end{mydef}

\subsubsection{Constructing feedback}
At the end of each execution, \testengine{} gathers attained coverage $cov$ together with fault detection statistics $stat$ to construct feedback.
These feedback information will be sent to \fuzzengine{} to guide input generation.

\subsection{Input Generation}
\label{subsec:input_gen}
\fuzzengine{} enables fault detection aware fuzzing with feedback information sent back from \testengine{}.
It updates interesting seeds queue $Q$ as well as failing inputs queue $F$ according to Algorithm \ref{alg:modified_cgf}. 
Inputs capable of killing mutants will be allocated more energy in order to generate more children from them (Algorithm \ref{alg:modified_mc}).
Note that \fuzzengine{} is a general component which can be replaced by any coverage guided fuzzer.

% \subsection{Bug Reproduction and Validation}
% \label{subsec:bug_triage_validate}
% \valengine{} executes save inputs to reproduce bugs and vulnerabilities found during fuzzing.
% For \textit{failing inputs}, \valengine{} executes them to make crash triage;
% For \textit{interesting inputs}, \valengine{} executes them with the differential testing framework proposed in Section \ref{subsubsec:mutation_killing_framework} to reveal and record potential bugs.
% We make \valengine{} work offline with all mutants in the mutant pool in order to adequately evaluate the fault detection capability for each saved input.
% Recall that we only use a small subset of mutants for mutation testing during fuzzing to ensure the efficiency of fuzzing.
% Executing inputs with all mutants can correct the underestimation of their fault detection capabilities caused by this trade-off.
% The recorded potential bugs are sent to developers for further validation. 

\section{EXPERIMENTAL SETUP}

% \textbf{\textit{Research Questions}}
In this paper, we evaluate how our approach influences the performance of CGF technique.  
In particular, we focus on the following research questions:
\begin{itemize}[leftmargin=*, topsep=3pt]
    \item \textbf{RQ1}: How mutation testing influences CGF in covering code?
    
    \item \textbf{RQ2}: How mutation testing influences CGF in terms of killing mutants?
    
    % \item \textbf{RQ3}: What mutants can be killed by each technique? 
    % \item \textbf{RQ3}: What impact can our approach make to CGF in terms of execution speed? 
    % \item \textbf{RQ1}: How effective is our approach in discovering bug-revealing inputs?
    % \item \textbf{RQ2}: How effective is our approach in finding vulnerabilities and functional bugs?
    % \item \textbf{RQ3}: What impact can our approach make to CGF in terms of code coverage and execution speed? 
\end{itemize}

% \begin{table*}[h]
%     \centering
%     \caption{Details of benchmarks}
%     \begin{tabular}{|l|l|l|l|l|l|l|}
%     \hline
%         PID & Project Name & Version & CID & Fuzz Case & \#Mutants & Criterion \\ \hline
%         P01 & commons-codec & 1.15 & C01 & Base64DecoderFuzz & 365 & ~ \\ \hline
%         ~ & ~ & ~ & C02 & Base64EncoderFuzz & 365 & ~ \\ \hline
%         ~ & ~ & ~ & C03 & CryptStringFuzz & 709 & ~ \\ \hline
%         P02 & commons-compress & 1.21 & C04 & ArchiveFuzz & 3614 & ~ \\ \hline
%         ~ & ~ & ~ & C05 & ChecksumCalculatingInputStreamFuzz & 16 & ~ \\ \hline
%         ~ & ~ & ~ & C06 & ZipLongFuzz & 83 & ~ \\ \hline
%         P03 & commons-lang & 3.12.0 & C07 & ArraySorterDoubleFuzz & 18 & ~ \\ \hline
%         ~ & ~ & ~ & C08 & ArraySorterStringFuzz & 18 & ~ \\ \hline
%         ~ & ~ & ~ & C09 & FractionDoubleFuzz & 8619 & ~ \\ \hline
%         ~ & ~ & ~ & C10 & FractionIntIntFuzz & 8619 & ~ \\ \hline
%         P04 & commons-math & 3.6.1 & C11 & DBSCANClustererFuzz & 7157 & ~ \\ \hline
%         ~ & ~ & ~ & C12 & LaguerreSolverFuzz & 8612 & ~ \\ \hline
%         P05 & maven & 3.8.4 & C13 & ExclusionArtifactFilterFuzz & 577 & ~ \\ \hline
%         ~ & ~ & ~ & C14 & MavenProjectFuzz & 1654 & ~ \\ \hline
%         ~ & ~ & ~ & C15 & ModelReaderFuzz & 3223 & ~ \\ \hline
%         ~ & ~ & ~ & C16 & ProjectSorterFuzz & 1654 & ~ \\ \hline
%     \end{tabular}
%     \label{tab:benchmarks}
% \end{table*}

% \textbf{\textit{Implementation.}}
\subsubsection*{Implementation}
We implement the proposed approach in Java.
Specifically, we implement fuzzing components with JQF \cite{jqf}, which is a widely used fuzzing framework for Java.
We implement mutant creation with PIT, which is the most widely used mutation testing engine for Java programs \cite{pitest}.
For test engine \testengine{}, we implement it using test runner supplied by Junit 4 and enable parallel execution with APIs supplied by JDK.
For on-the-fly mutant selection, we implement it as an interface \texttt{MutantSelectionStrategy}.
For differential outputs checking, we implement it as an interface \texttt{Criterion}. 
\texttt{MutantSelectionStrategy} can be extended to build subtypes to implement different mutant selection strategy.
Similarly, \texttt{Criterion} can be extended to meet special needs of checking different kinds of outputs.
To conduct experiments, we implement \texttt{BasicRandomStrategy} to select fixed number of mutants from mutant pool during fuzzing.
For PUT with serializable output, we build a subtype, \texttt{SerializableCriterion}, to serialize the outputs of the PUT and its mutants and compare the bytes after serialization.

% \textbf{\textit{Fuzz Cases.}}
% \subsubsection*{Fuzz Cases}
\subsubsection*{Benchmarks}

% \begin{table}[htb]
% \caption{Details of Benchmarks.}
% \label{tab:fuzz_cases}
% \begin{tabular}{|l|l|l|l|}
% \hline
% \textbf{CID} & \textbf{Fuzz Case} & \textbf{\#Mutants} & \textbf{LoC of Bench.}\\ \hline
% C01 & SortingFuzz          &  69       &  88     \\ \hline
% C02 & MatrixInverseFuzz    &  75       &  65     \\ \hline
% C03 & SuffixArrayFuzz      &  215      &  219    \\ \hline
% C04 & SimpleRegressionFuzz &  49783    &  208891 \\ \hline
% C05 & DivFuzz              &  577      &  30396  \\ \hline
% \end{tabular}
% \end{table}

The details of the benchmarks are shown at Table \ref{tab:fuzz_cases}.
All our benchmarks are from Github\cite{github} or previous works.
One of our authors manually build \nFuzzCase{} fuzz cases for \nBenchmark{} benchmarks.
In particular, C01, C02 and C03 are experimental objects used in \cite{sun2019maf}; C04 and C05 are widely adopted open-sourced subjects.
The version of Apache Commons Math3 3.6.1 and Apache Commons Numbers is 1.0.
Column \textbf{\#Mutants} illustrates the total number of mutants we created for each benchmark using PIT default mutators\footnote{\url{https://pitest.org/quickstart/mutators/}}, and the last column shows the number of non-blank lines of code.

\begin{table}[htb]
\caption{Details of Benchmarks.}
\label{tab:fuzz_cases}
\begin{tabular}{|l|l|l|l|}
\hline
\textbf{CID} & \textbf{Fuzz Case} & \textbf{\#Mutants} & \textbf{LoC of Bench.}\\ \hline
C01 & SortingFuzz          &  69       &  88     \\ \hline
C02 & MatrixInverseFuzz    &  75       &  65     \\ \hline
C03 & SuffixArrayFuzz      &  215      &  219    \\ \hline
C04 & SimpleRegressionFuzz &  49783    &  208891 \\ \hline
C05 & DivFuzz              &  577      &  30396  \\ \hline
\end{tabular}
\end{table}

% The numbers of mutants we created for fuzz cases are illustrated at column \textbf{\#Mutants}.

% \textbf{\textit{Techniques.}}
\subsubsection*{Techniques} 
We choose state-of-the-art CGF technique \zest{} \cite{zest} as baseline.
\zest{} is the only fuzzing technique for Java to our best knowledge.
We modified \zest{} with our approach from two aspects, and construct two techniques, namely \nmuzest{} and \pmuzest{}.
``N'' and ``P'' indicate how we enhance fuzzing campaigns with mutation testing:
for \nmuzest{} we influences fuzzing with mutation testing ``negatively'', that is,  
we punish the inputs that kill mutants which have been killed in previous by lessening children inputs generated from them;
 \cite{chan1996proportional}.
for \pmuzest{} we influences fuzzing with mutation testing ``positively'', that is,
we reward the inputs which could kill mutants (no matter whether the mutants have been killed in previous).
\nmuzest{} and \pmuzest{} guide CGF in very different ways:
with \nmuzest{} we guide fuzzing to explore space far away from the mutants that have been detected in previous, which follows the observation of \cite{chan1996proportional};
with \pmuzest{}, on the contrary, we want to guide fuzzing towards exploring deeper paths. 

We instantiate Algorithm \ref{alg:modified_mc} differently to implement \nmuzest{} and \pmuzest{}.
We choose \mfactor{} (the multiplication factor for number of children to produce for favored inputs used by \texttt{ZestGuidance}) as the modifier factor used in Algorithm \ref{alg:modified_mc}.
In particular, for \nmuzest{} we 
(1) divide the number of generated children inputs with \mfactor{} when inputs kill old mutants, and
(2) multiply the number of generated children kill with \mfactor{} when inputs detect fresh mutants;
for \pmuzest{}, we just multiply the number of generated children inputs with \mfactor{} when inputs kill any mutants.

% \textbf{\textit{Experiments.}}
\subsubsection*{Experiments}
We combine fuzz cases with techniques to construct \nFuzzPair{} fuzz campaigns (each campaign is a pair \fuzzPair{}).
We run each campaign for \duration{} following the setup of \cite{zest} to obtain seed corpus.
To remove the impact of randomness, we repeat each campaign for \nRepeat{} times.
We use \texttt{BasicRandomStategy} to select \nSelect{} mutants during each campaign and use \nThread{} threads (\nSelect{} for mutants and 1 for original PUT) to run these mutants along with original PUT concurrently.
We set \textit{extra execution time ratio} as 0.1, which means the time upper bound of executing a mutant is 10\% more than executing the original PUT. 
After finishing all fuzzing campaigns, we run scripts to reproduce each preserved inputs to compute average branch coverage and mutant killing rate obtained in 10 runs. 
All experiments were run on a cloud machine with 32GB RAM and 16-core Intel Core Processor CPU.

\section{Evaluation}

% To answer \textbf{RQ1}, we reproduce every saved inputs and identify bug-revealing ones with \valengine{} to calculate \textit{bug-revealing portion} (BRP) as $BRP = \frac{\#I_{bug}}{\#I}$ where $\#I_{bug}$ denotes the number of bug-revealing inputs, $\#I$ the number of total saved inputs.
% To answer \textbf{RQ2}, we adopt the number of unique crashes (NUC) to indicate the effectiveness of finding vulnerabilities.
% Note that NUC is used widely in fuzzing researches \cite{zest, perffuzz} \qrx{more citations?}.
% Moreover, we use a similar metric, namely the number of unique bugs (NUB) to stand for the effectiveness of finding function bugs.
% To answer \textbf{RQ3}, we use branch coverage and the number of executions.
% We run each fuzz case for \duration{}.
% To reduce the impact of randomness, we repeat each fuzz case for \nRepeat{} times.

\begin{figure*}[htbp]
\centering
\begin{subfigure}{0.19\textwidth}
    \includegraphics[width=\textwidth]{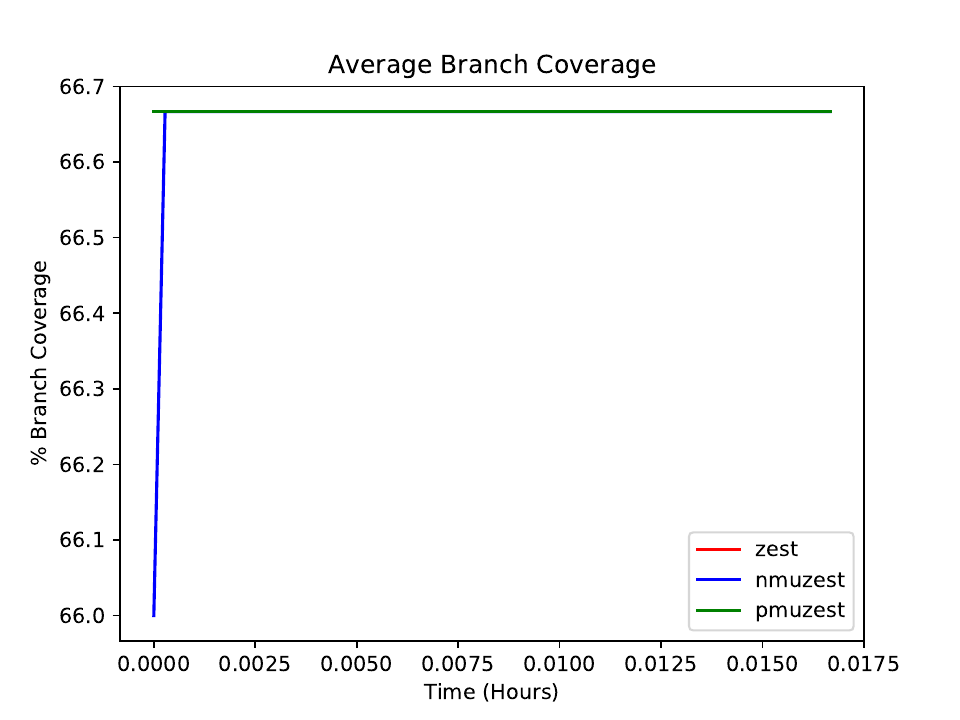}
    \caption{C01-1min}
    \label{subfig:bC01-1min}
\end{subfigure}
\hfill
\begin{subfigure}{0.19\textwidth}
    \includegraphics[width=\textwidth]{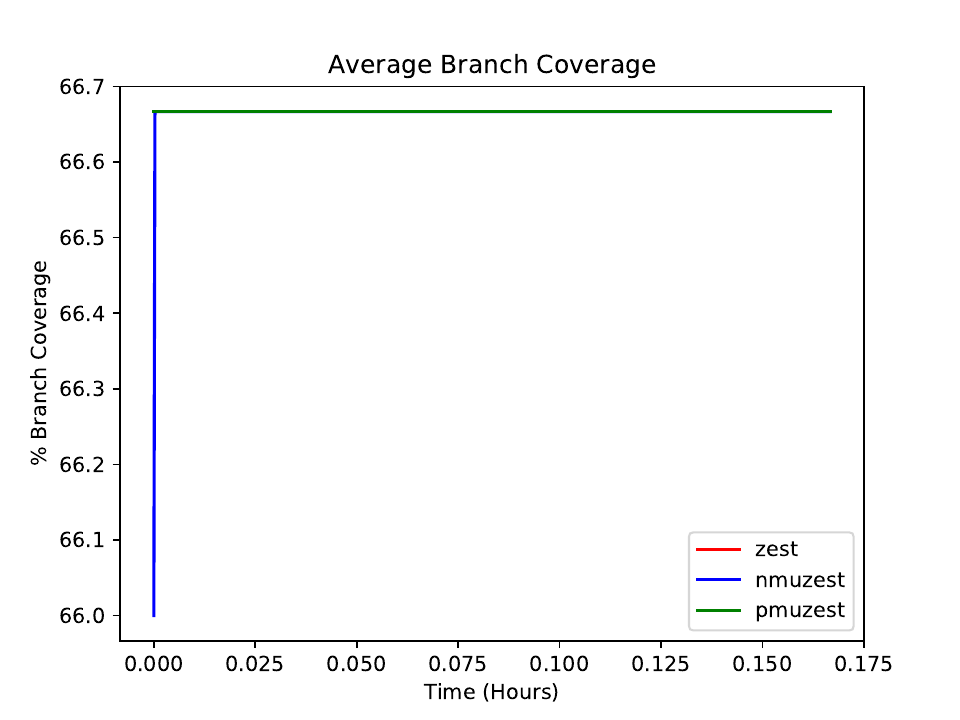}
    \caption{C01-10min}
    \label{subfig:bC01-10min}
\end{subfigure}
\hfill
\begin{subfigure}{0.19\textwidth}
    \includegraphics[width=\textwidth]{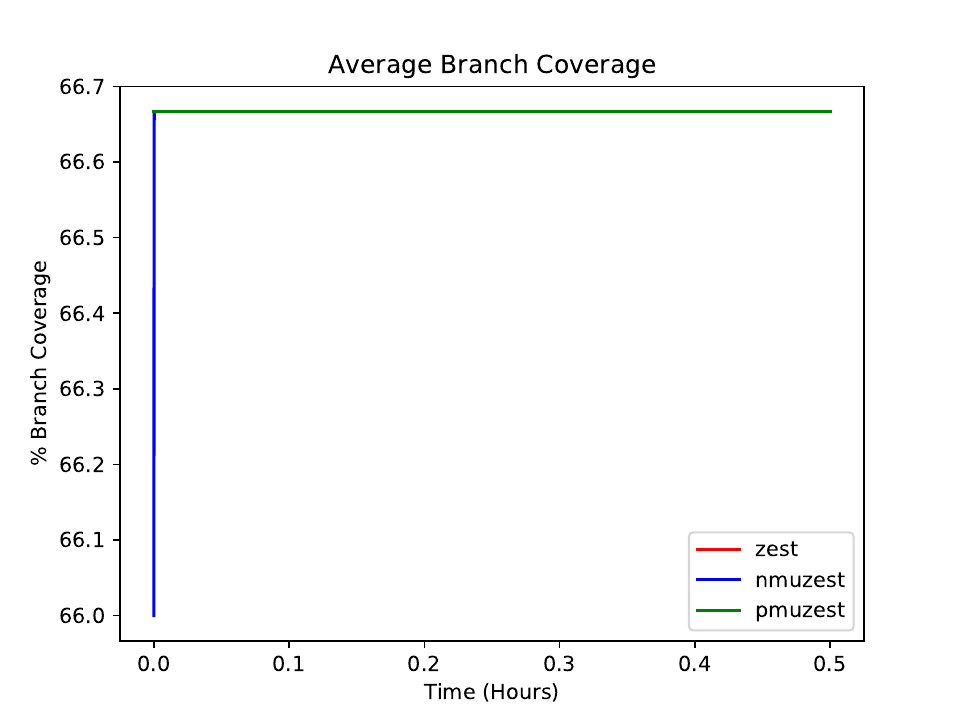}
    \caption{C01-30min}
    \label{subfig:bC01-30min}
\end{subfigure}
\hfill
\begin{subfigure}{0.19\textwidth}
    \includegraphics[width=\textwidth]{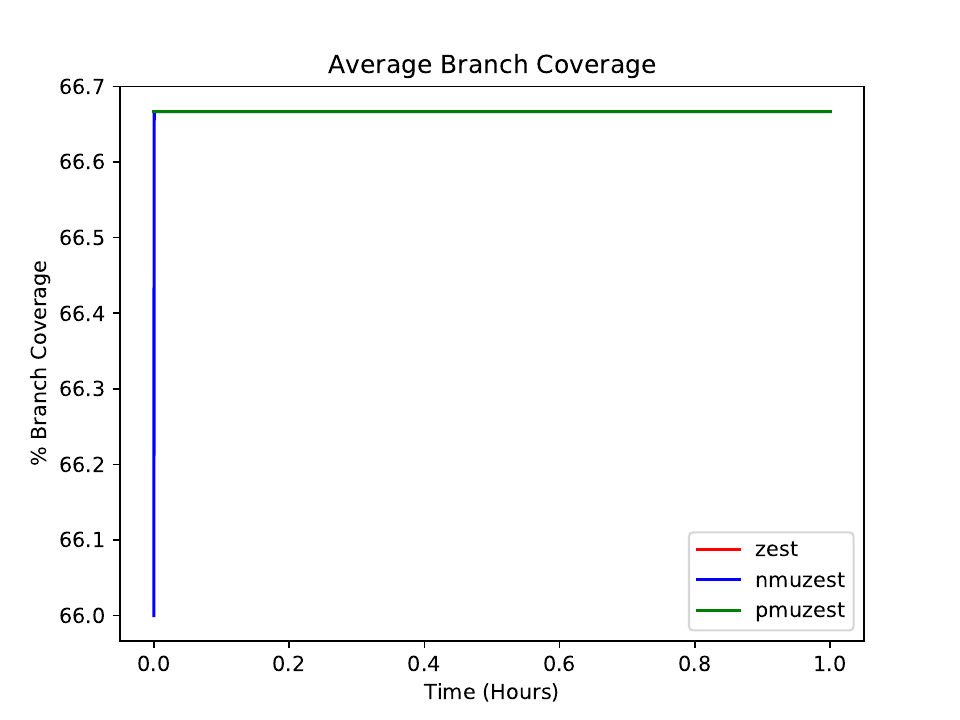}
    \caption{C01-1hour}
    \label{subfig:bC01-1hour}
\end{subfigure}
\hfill
\begin{subfigure}{0.19\textwidth}
    \includegraphics[width=\textwidth]{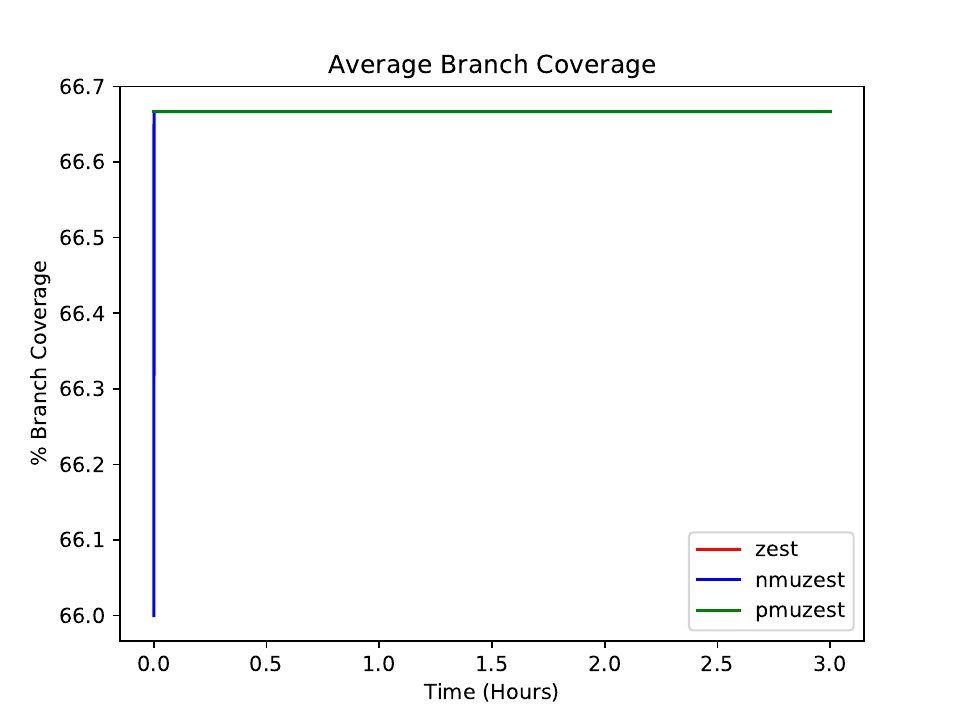}
    \caption{C01-3hour}
    \label{subfig:bC01-3hour}
\end{subfigure}
% -------------------------------------- %
\begin{subfigure}{0.19\textwidth}
    \includegraphics[width=\textwidth]{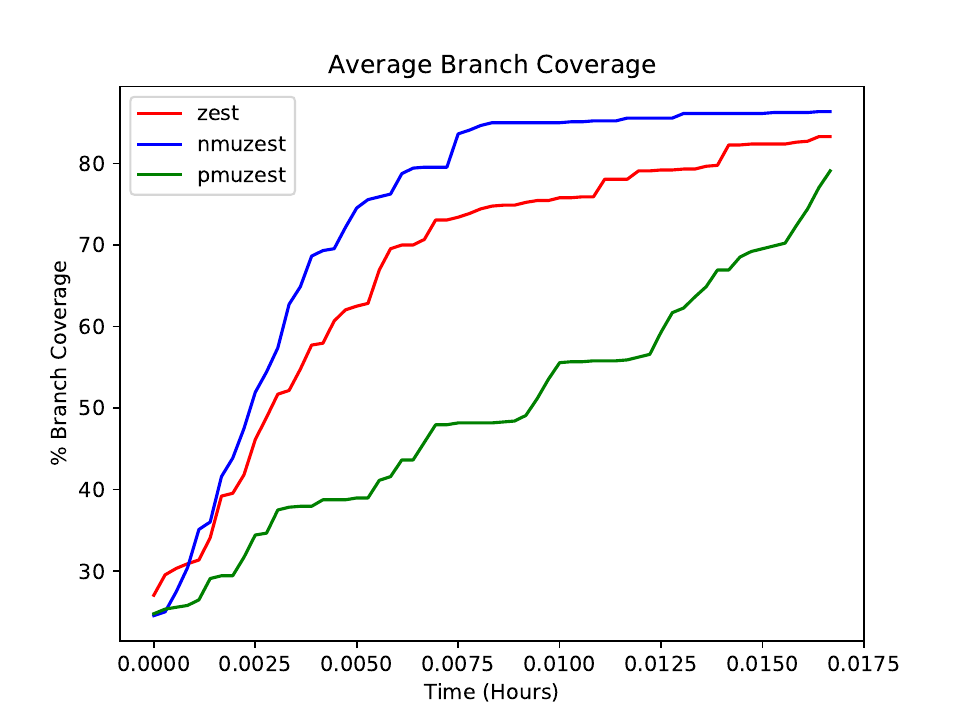}
    \caption{C02-1min}
    \label{subfig:bC02-1min}
\end{subfigure}
\hfill
\begin{subfigure}{0.19\textwidth}
    \includegraphics[width=\textwidth]{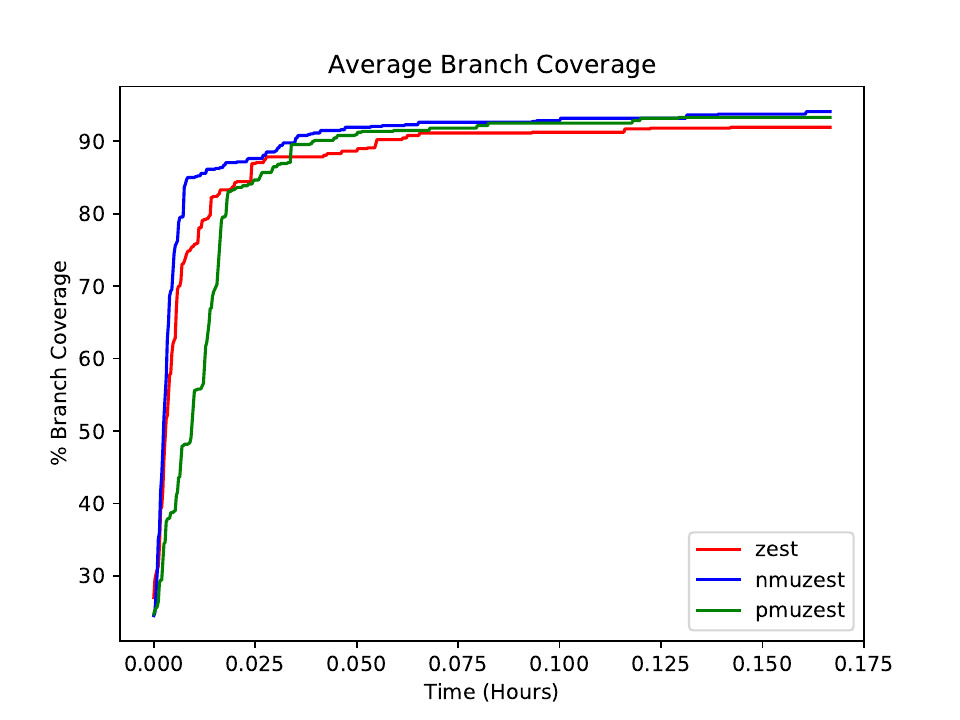}
    \caption{C02-10min}
    \label{subfig:bC02-10min}
\end{subfigure}
\hfill
\begin{subfigure}{0.19\textwidth}
    \includegraphics[width=\textwidth]{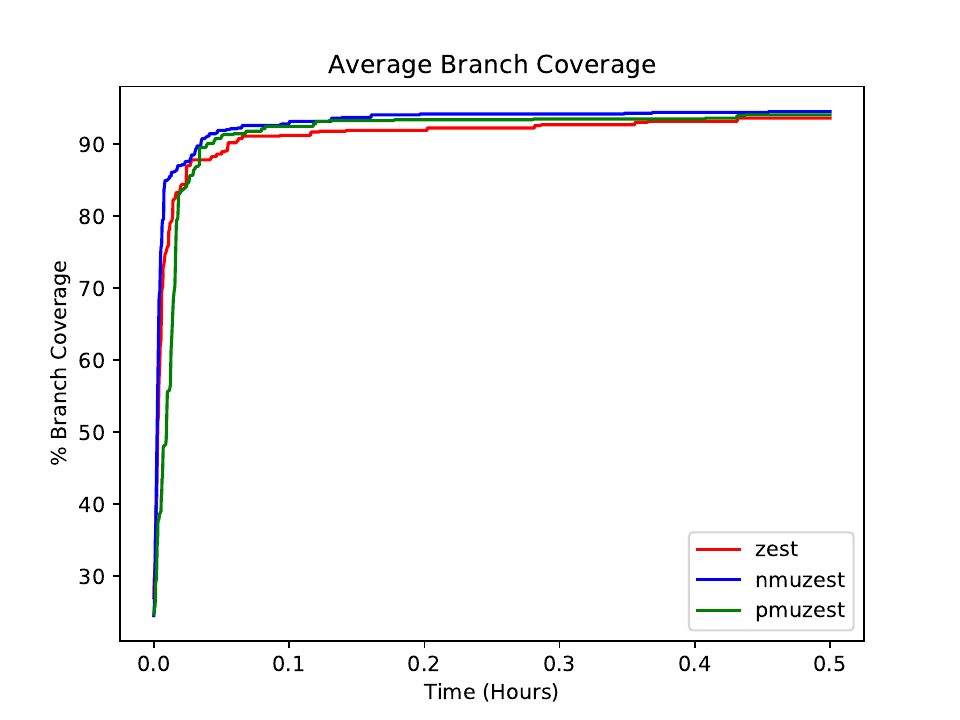}
    \caption{C02-30min}
    \label{subfig:bC02-30min}
\end{subfigure}
\hfill
\begin{subfigure}{0.19\textwidth}
    \includegraphics[width=\textwidth]{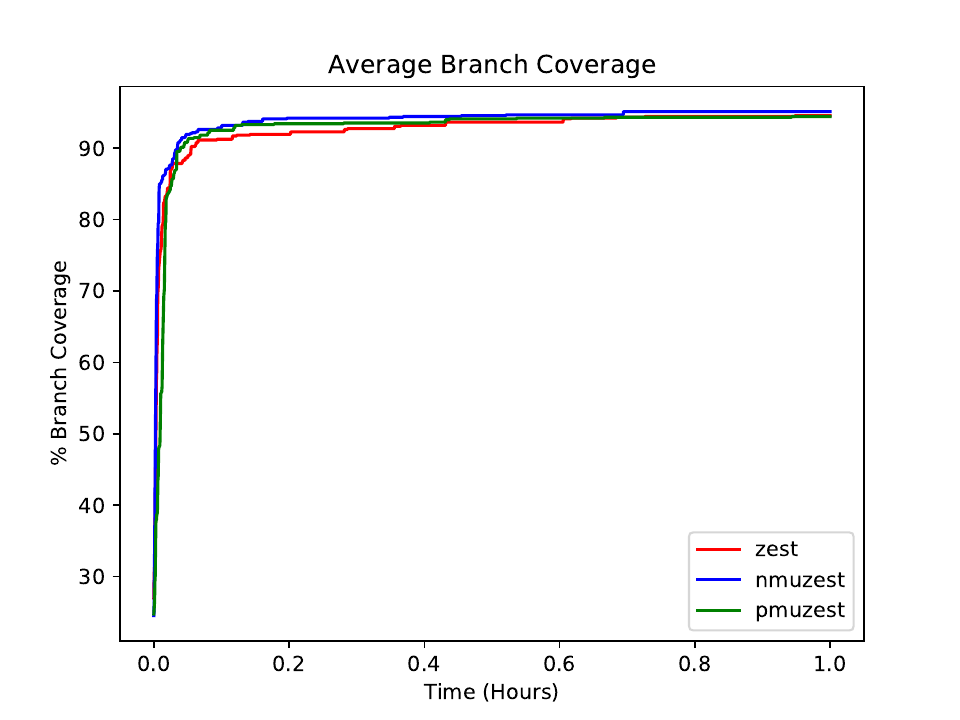}
    \caption{C02-1hour}
    \label{subfig:bC02-1hour}
\end{subfigure}
\hfill
\begin{subfigure}{0.19\textwidth}
    \includegraphics[width=\textwidth]{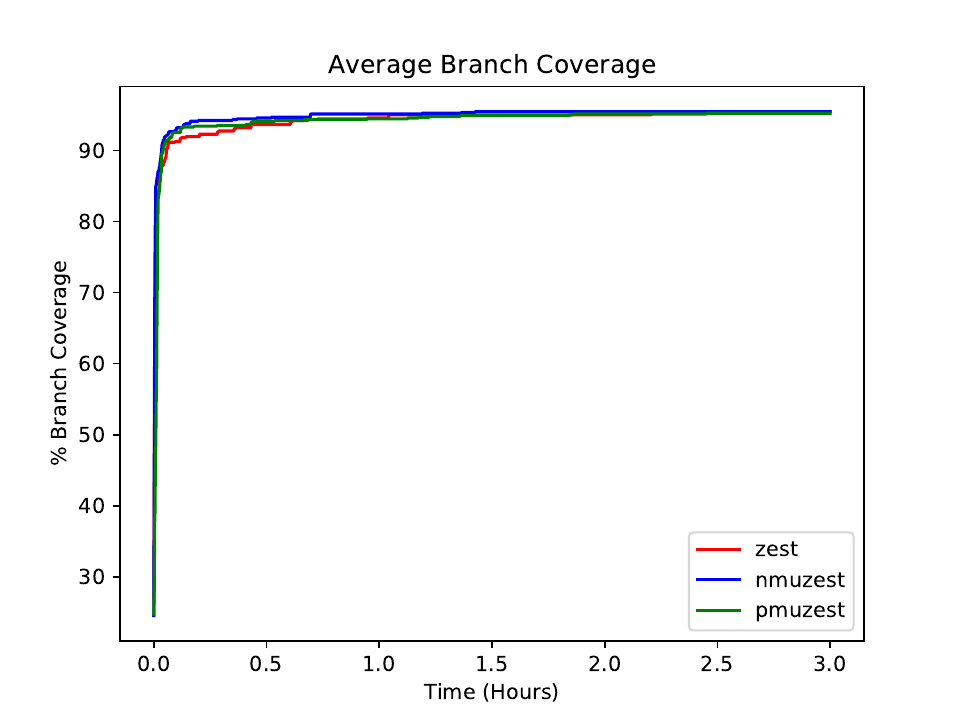}
    \caption{C02-3hour}
    \label{subfig:bC02-3hour}
\end{subfigure}
% -------------------------------------- %
\begin{subfigure}{0.19\textwidth}
    \includegraphics[width=\textwidth]{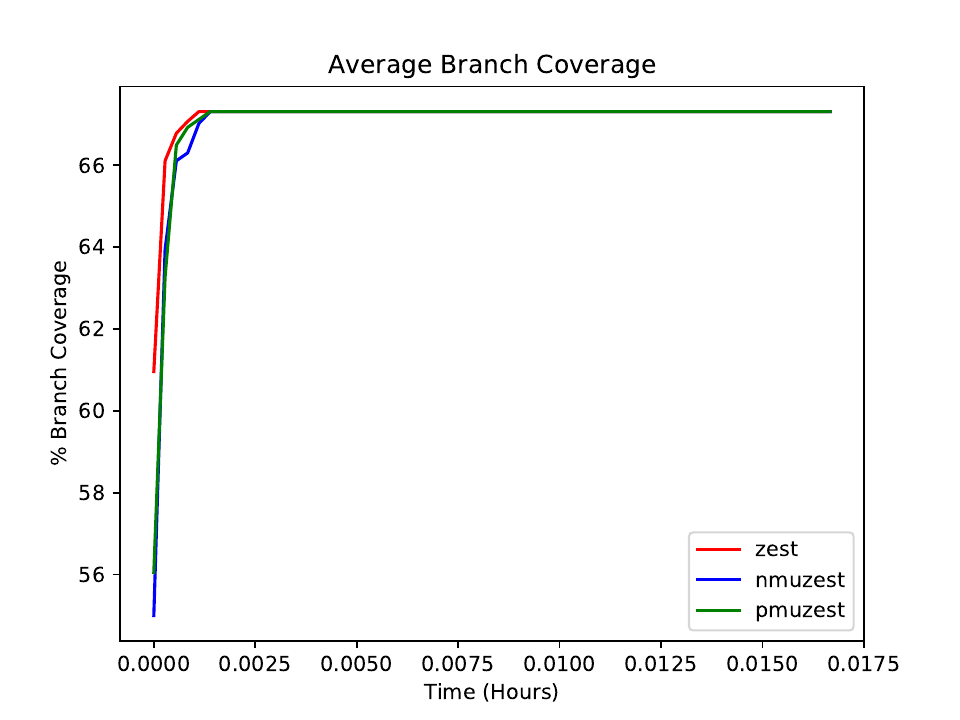}
    \caption{C03-1min}
    \label{subfig:bC03-1min}
\end{subfigure}
\hfill
\begin{subfigure}{0.19\textwidth}
    \includegraphics[width=\textwidth]{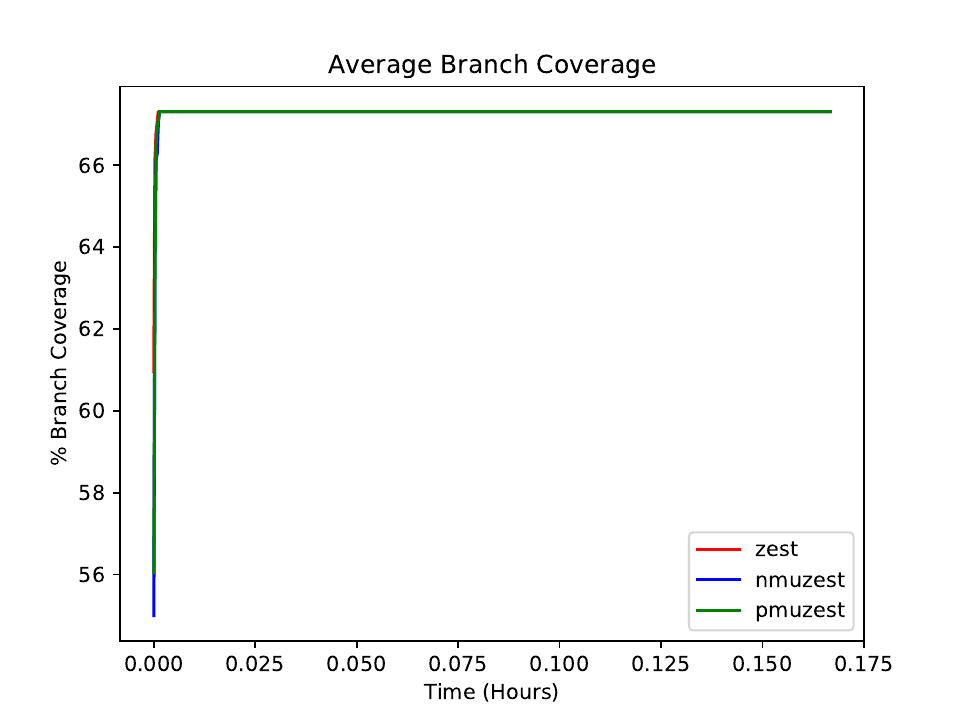}
    \caption{C03-10min}
    \label{subfig:bC03-10min}
\end{subfigure}
\hfill
\begin{subfigure}{0.19\textwidth}
    \includegraphics[width=\textwidth]{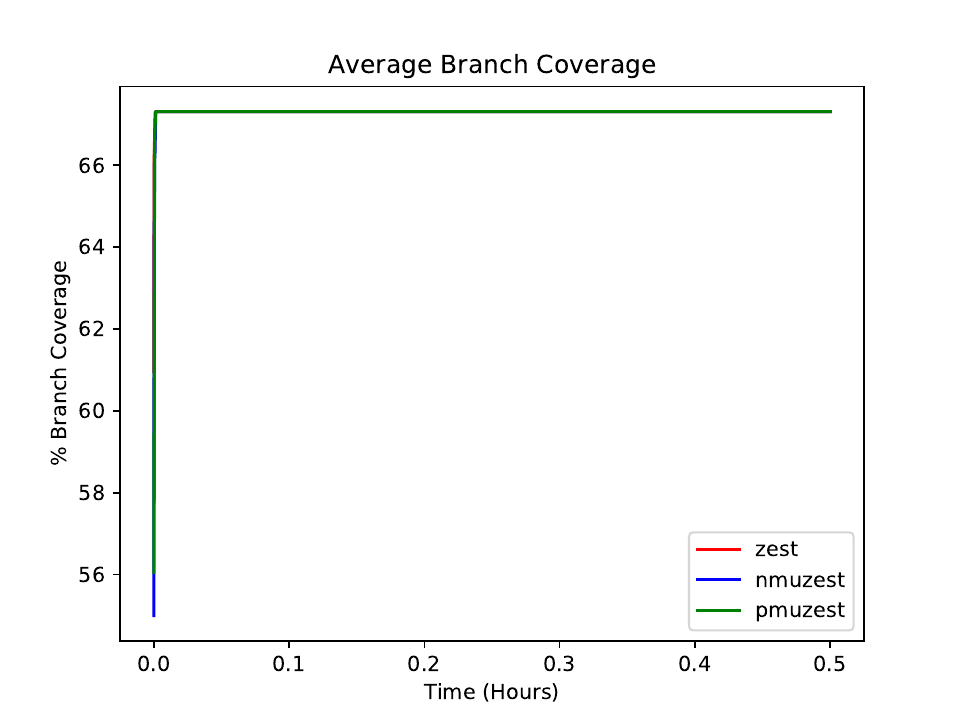}
    \caption{C03-30min}
    \label{subfig:bC03-30min}
\end{subfigure}
\hfill
\begin{subfigure}{0.19\textwidth}
    \includegraphics[width=\textwidth]{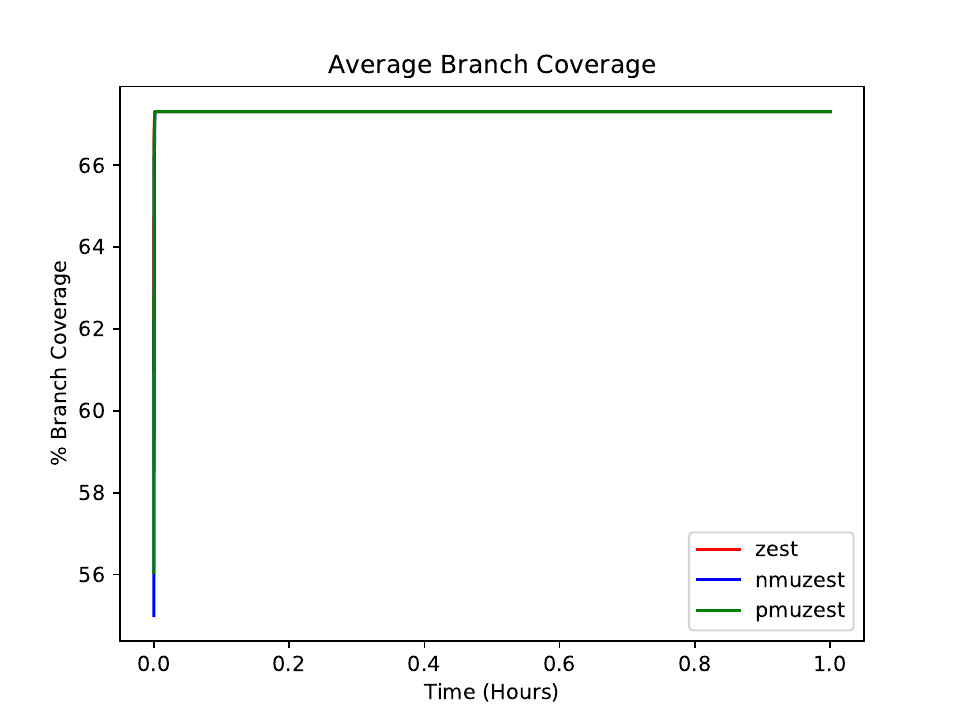}
    \caption{C03-1hour}
    \label{subfig:bC03-1hour}
\end{subfigure}
\hfill
\begin{subfigure}{0.19\textwidth}
    \includegraphics[width=\textwidth]{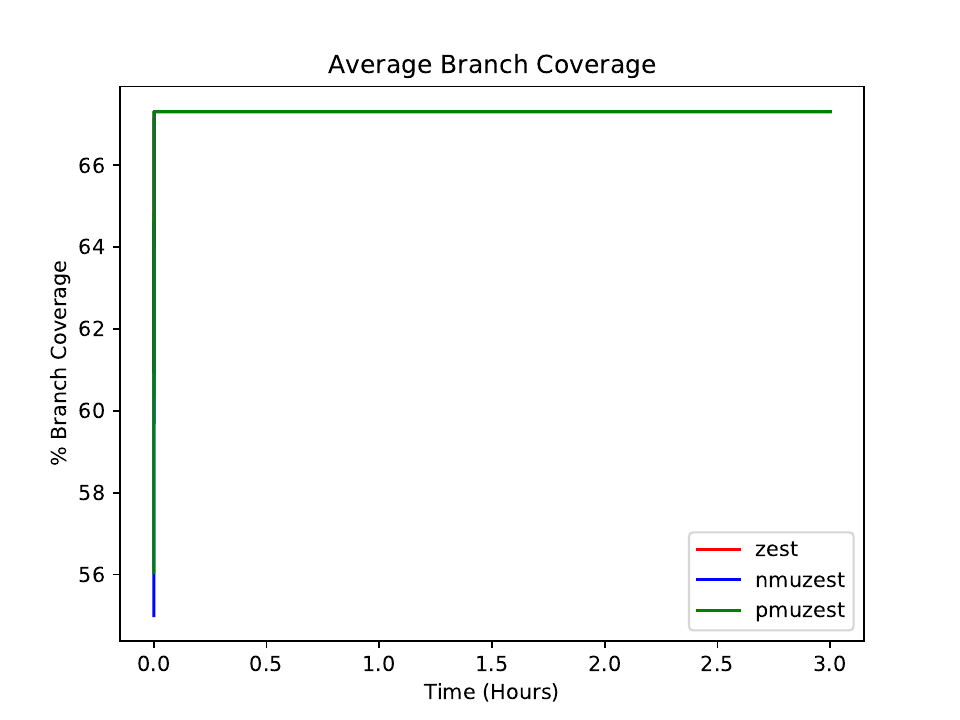}
    \caption{C03-3hour}
    \label{subfig:bC03-1hour}
\end{subfigure}
% -------------------------------------- %
\begin{subfigure}{0.19\textwidth}
    \includegraphics[width=\textwidth]{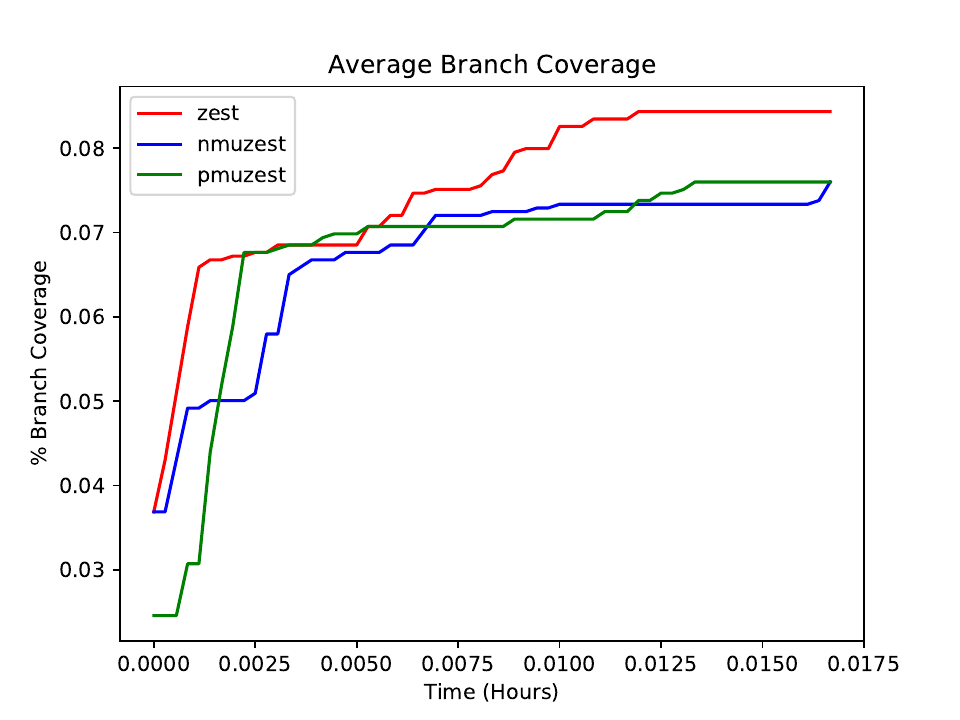}
    \caption{C04-1min}
    \label{subfig:bC04-1min}
\end{subfigure}
\hfill
\begin{subfigure}{0.19\textwidth}
    \includegraphics[width=\textwidth]{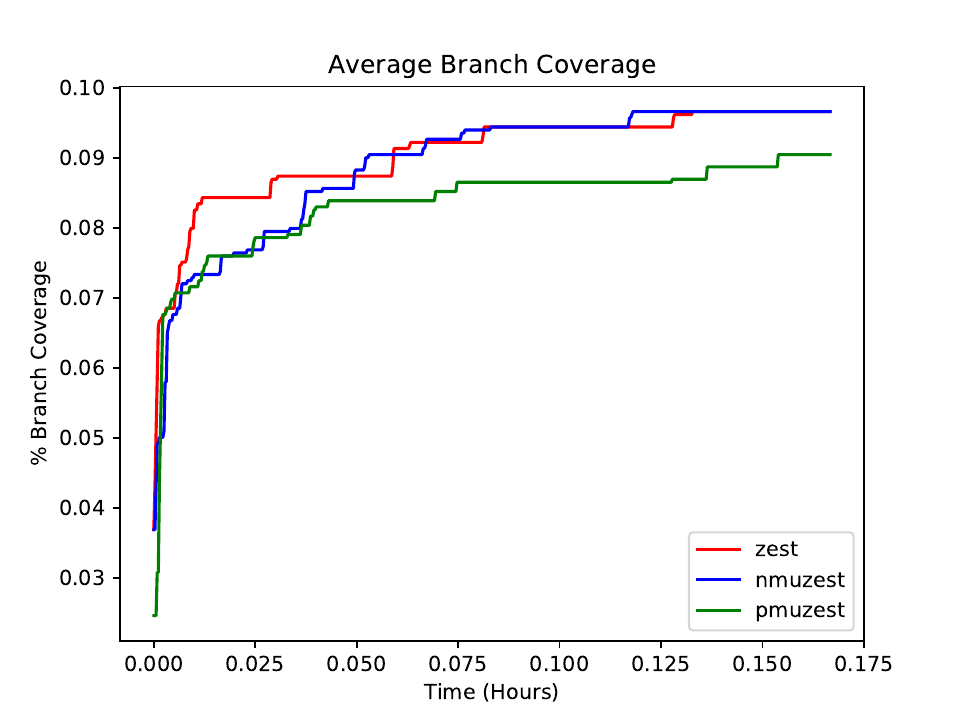}
    \caption{C04-10min}
    \label{subfig:bC04-10min}
\end{subfigure}
\hfill
\begin{subfigure}{0.19\textwidth}
    \includegraphics[width=\textwidth]{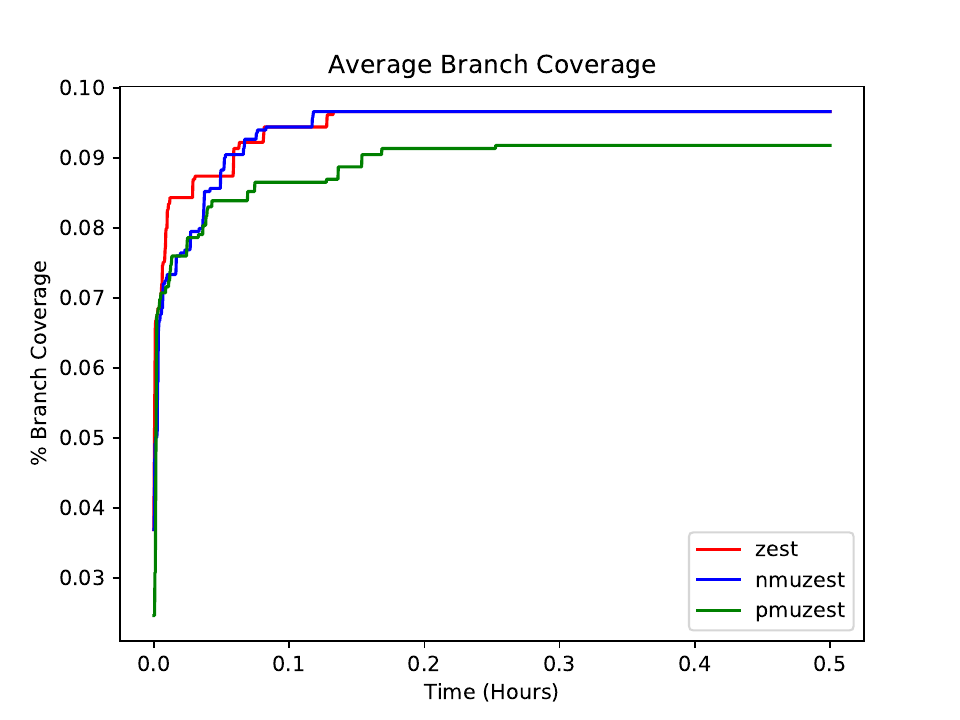}
    \caption{C04-30min}
    \label{subfig:bC04-30min}
\end{subfigure}
\hfill
\begin{subfigure}{0.19\textwidth}
    \includegraphics[width=\textwidth]{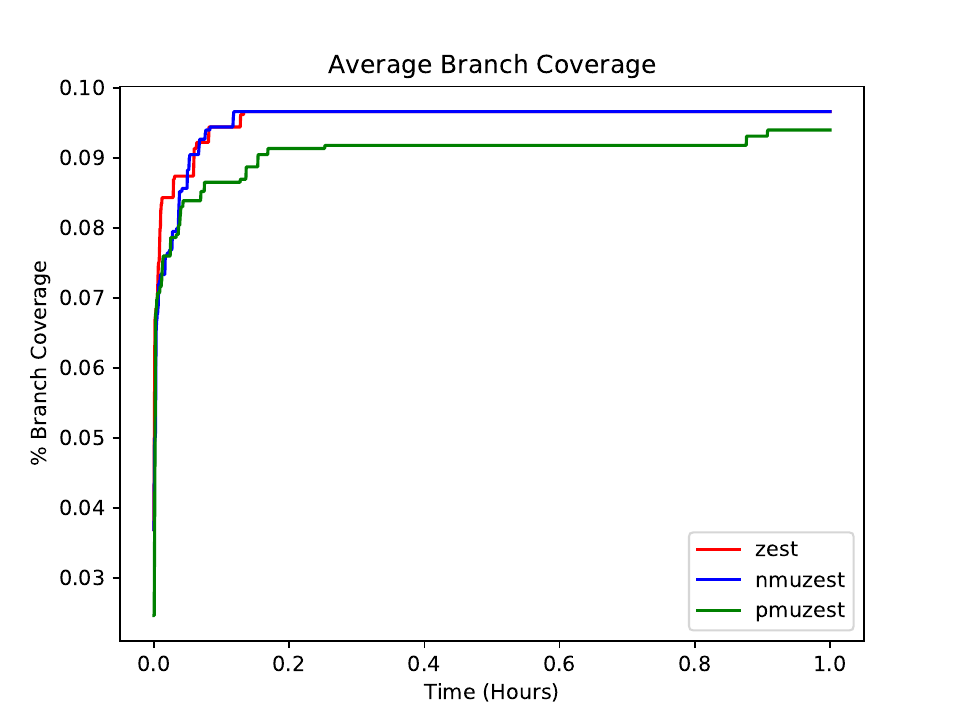}
    \caption{C04-1hour}
    \label{subfig:bC04-1hour}
\end{subfigure}
\hfill
\begin{subfigure}{0.19\textwidth}
    \includegraphics[width=\textwidth]{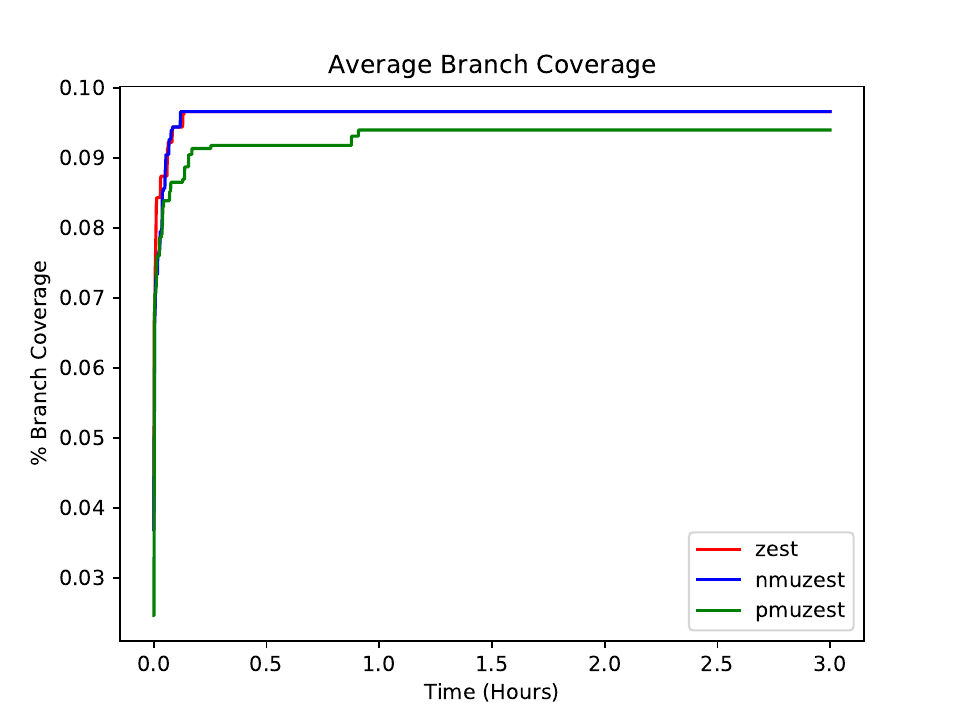}
    \caption{C04-3hour}
    \label{subfig:bC04-3hour}
\end{subfigure}
% -------------------------------------- %
\begin{subfigure}{0.19\textwidth}
    \includegraphics[width=\textwidth]{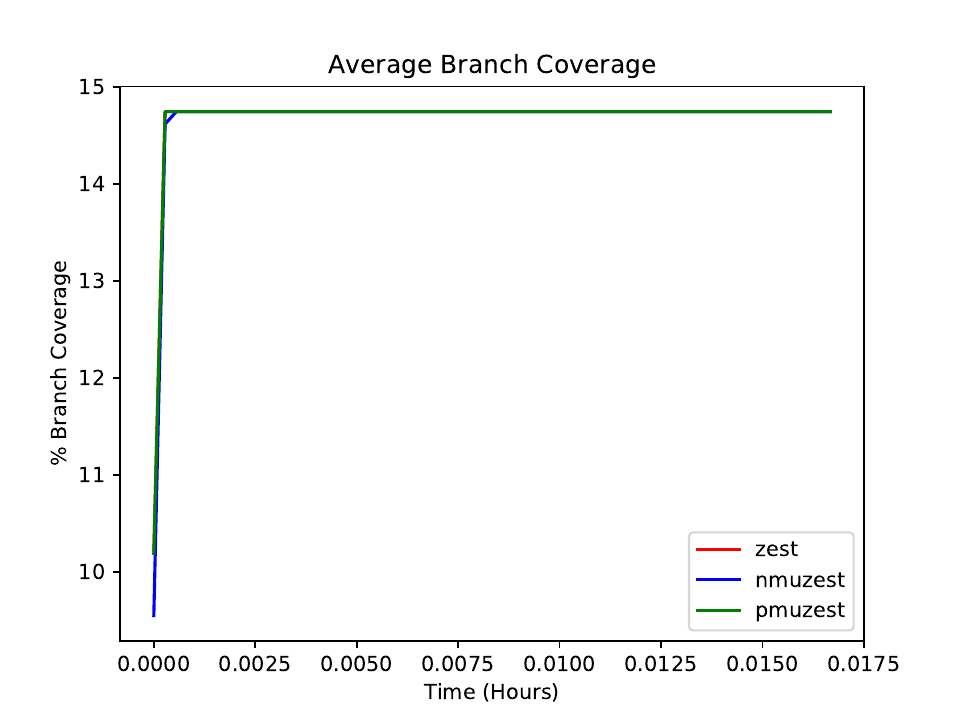}
    \caption{C05-1min}
    \label{subfig:bC05-1min}
\end{subfigure}
\hfill
\begin{subfigure}{0.19\textwidth}
    \includegraphics[width=\textwidth]{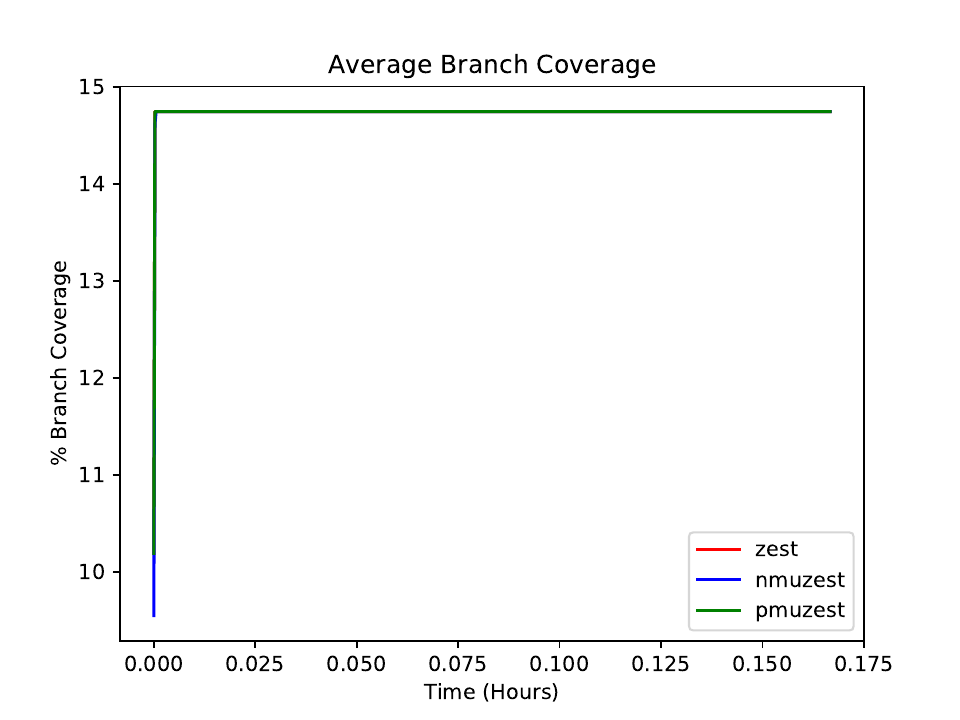}
    \caption{C05-10min}
    \label{subfig:bC05-10min}
\end{subfigure}
\hfill
\begin{subfigure}{0.19\textwidth}
    \includegraphics[width=\textwidth]{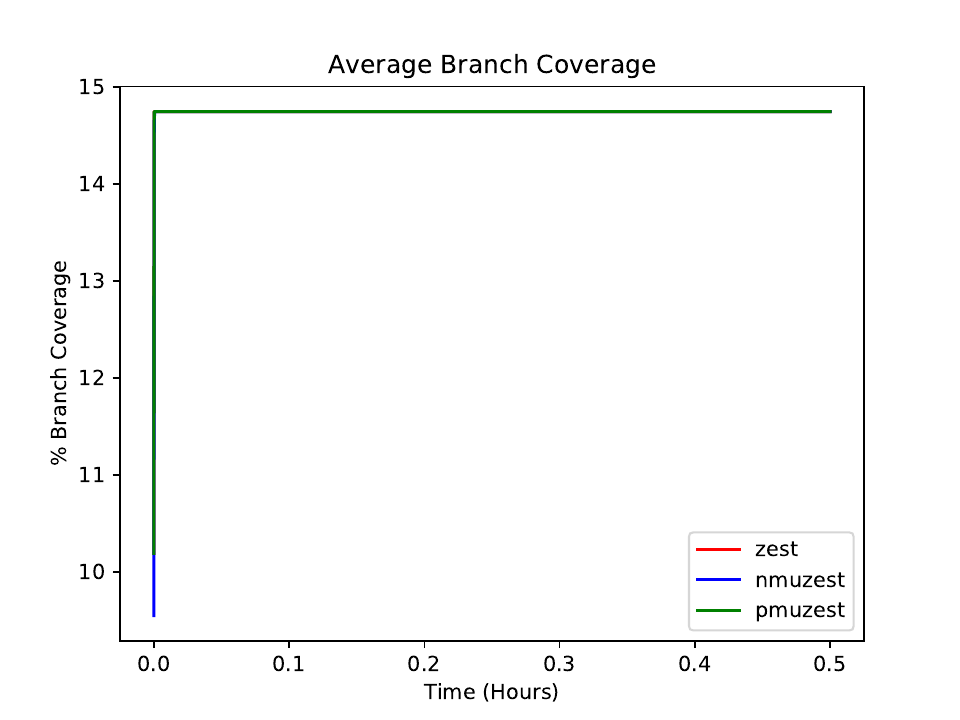}
    \caption{C05-30min}
    \label{subfig:bC05-30min}
\end{subfigure}
\hfill
\begin{subfigure}{0.19\textwidth}
    \includegraphics[width=\textwidth]{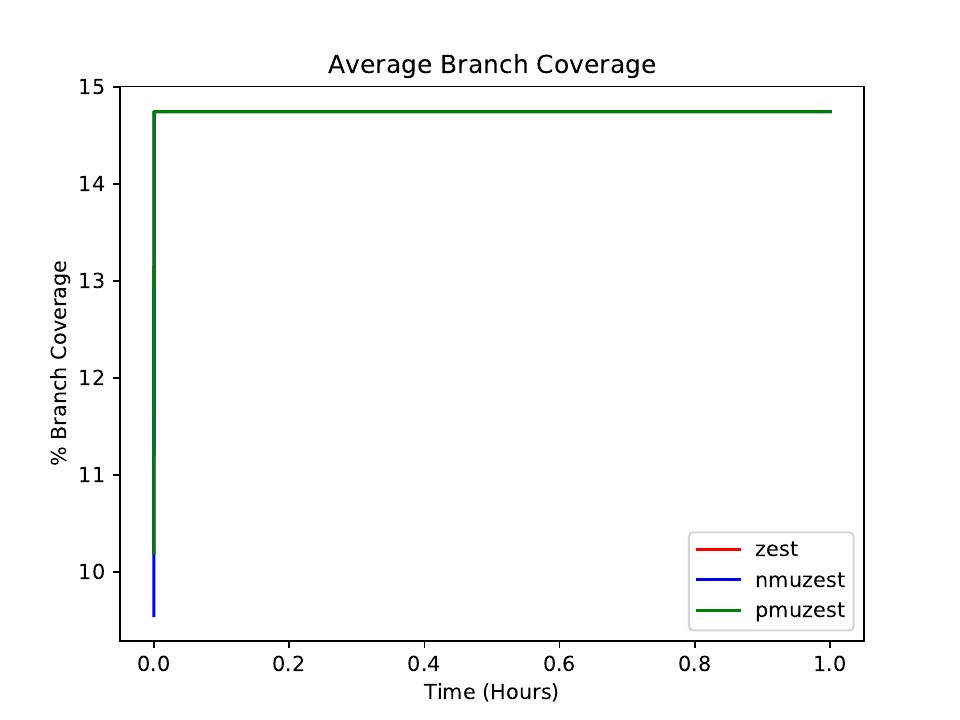}
    \caption{C05-1hour}
    \label{subfig:bC05-1hour}
\end{subfigure}
\hfill
\begin{subfigure}{0.19\textwidth}
    \includegraphics[width=\textwidth]{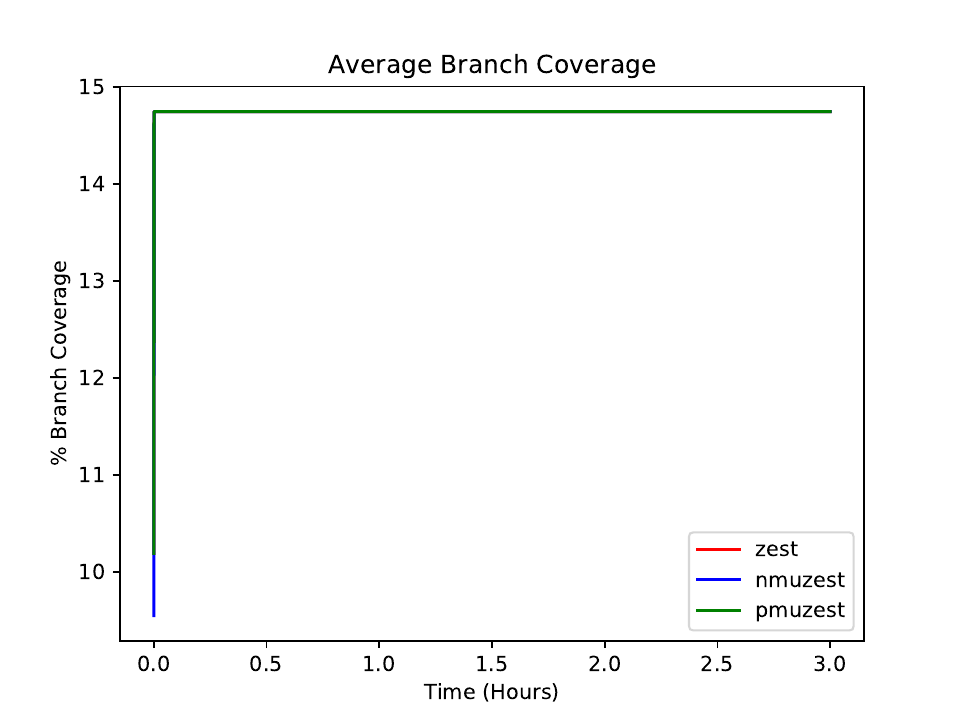}
    \caption{C05-3hour}
    \label{subfig:bC05-3hour}
\end{subfigure}
% -------------------------------------- %
\caption{
Average branch coverage rates for different fuzz campaigns in different time durations.
The x-axis of each figure represents time points and the y-axis represents the achieved branch coverage rates.
% The curves for \zest{}, \nmuzest{} and \pmuzest{} are colored in red, blue and green respectively.
}
\label{fig:branch_covs}
\end{figure*}

% \subsection{RQ1: Effectiveness in Covering Code}
\subsection{RQ1: Effectiveness in Code Coverage}
To illustrate the performances of different techniques in covering code, we reproduce the seed inputs created during fuzzing campaigns and attach the most recent time points to them.
For example, suppose that a certain fuzz campaign generates 4 input seeds \texttt{id\_0000}$\sim$\texttt{id\_0003} during 0$\sim$1s, then we attach the average coverage achieve by \texttt{id\_0000}$\sim$\texttt{id\_0003} to time point 1s.
Figure \ref{fig:branch_covs} shows the average branch coverage achieved during different fuzz campaigns.
Each row of sub-figures represents the changing tendencies in different time durations from \textbf{1 minute} (1m) to \textbf{3 hours} (3h).
We use red, blue and green lines to illustrate coverage curves produced by \zest{}, \pmuzest{} and \nmuzest{} respectively.
The x-axis represents time points whereas y-axis the achieved average branch coverage rates.
We obtain the following observations:

\vspace{-1mm}
\begin{itemize}[leftmargin=*, topsep=3pt]
    \item \textit{\textbf{Compare among durations.}}
    Branch coverage increases drastically at the beginning of each fuzz campaigns, especially in the first minute (the first column of Figure \ref{fig:branch_covs}).
    
    \item \textit{\textbf{Compare among techniques.}}
    In terms of covering code, \nmuzest{} performs slightly better then \zest{}, while \pmuzest{} performs slightly worse.
    In C01, C03, C05, the maximum coverage rates achieved by each techniques are nearly the same and the curves appear to overlap with each other.
    In C02, although the maximum coverage rates achieved by each technique are nearly the same, the increasing tendency of \nmuzest{} is much more drastic than that of \zest{}. 
    By contrast, the increasing tendency of \pmuzest{} is much gentler.
    In C04, although the increasing tendency of \zest{} is the most drastic compared to \nmuzest{} and \pmuzest{} in the first minute of campaign, it is quickly caught up by \pmuzest{} in the next few minutes.
    What's more, the maximum coverage rate achieved by \zest{} is also worse than that of \nmuzest{}.
    
    \item \textit{\textbf{Compare among cases.}}
    The performance in covering codes varies for different fuzz cases, appears as two aspects:
    (1) \textbf{Increasing tendencies.}
    In C01, C05 and C05, coverage curves overlap with each other and the maximum coverage rates are similar.
    Code coverage rates in these cases rise rapidly in the first minute to achieve maximum and become flat in the rest of campaigns.
    On the contrary, coverage curves in C04 and C04 keep rising for more than 30 minutes.   
    (2) \textbf{Maximum coverage rates.}
    In C01$\sim$C03, the maximum coverage rates are more than 60\%, whereas the maximum coverage rates in C04 and C05 are far less (below 15\%), especially in C04, which is no more than 0.10\% (0.096\%).
\end{itemize}

\begin{figure*}[htbp]
\centering
\begin{subfigure}{0.19\textwidth}
    \includegraphics[width=\textwidth]{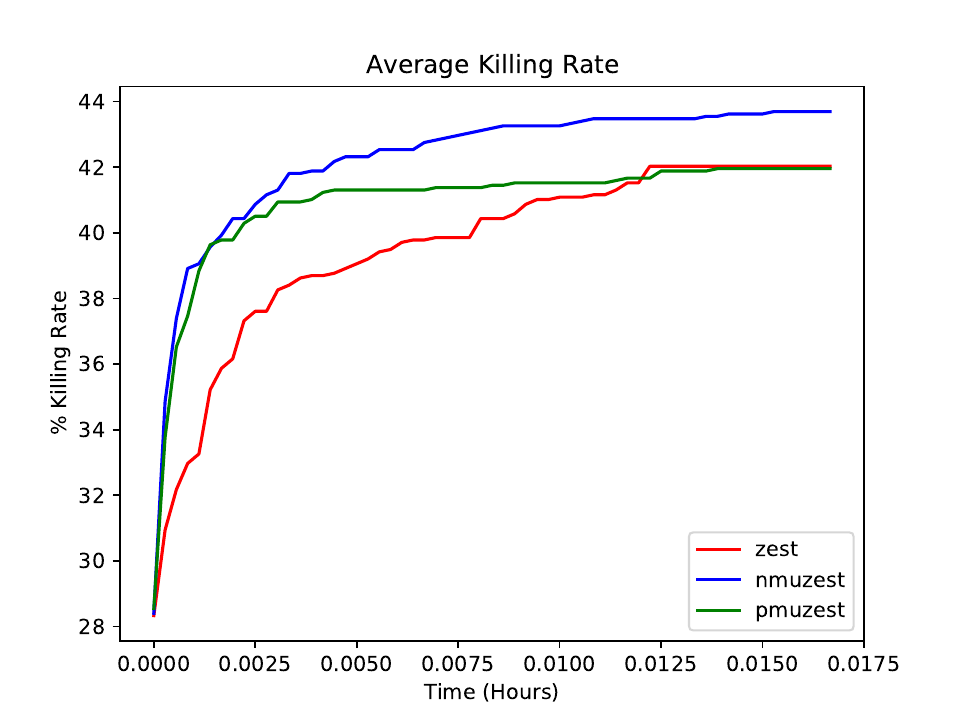}
    \caption{C01-1min}
    \label{subfig:mC01-1min}
\end{subfigure}
\hfill
\begin{subfigure}{0.19\textwidth}
    \includegraphics[width=\textwidth]{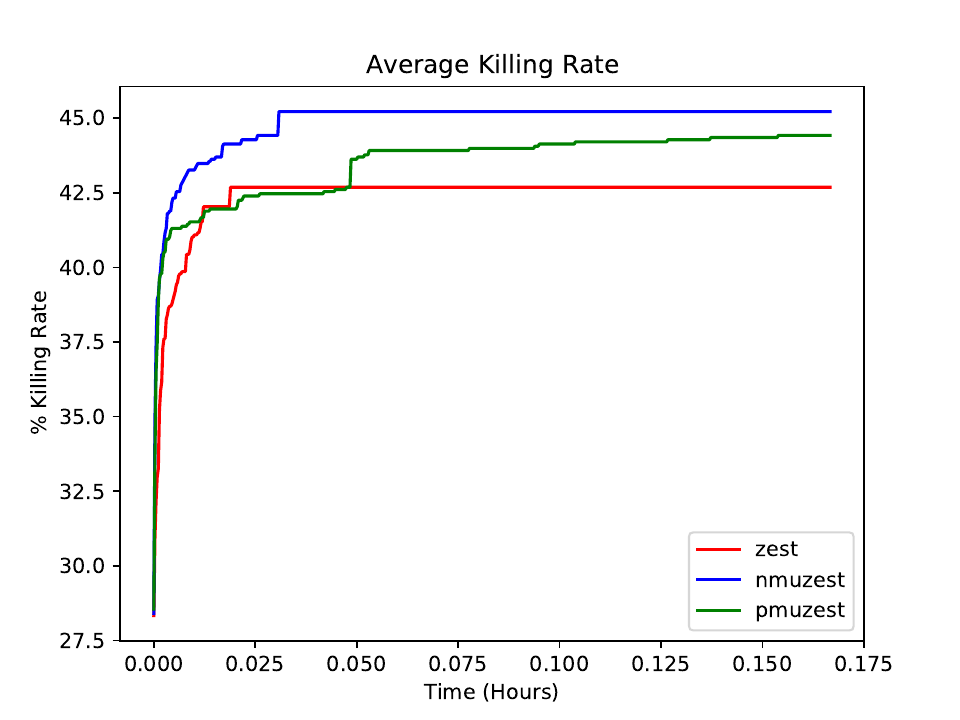}
    \caption{C01-10min}
    \label{subfig:mC01-10min}
\end{subfigure}
\hfill
\begin{subfigure}{0.19\textwidth}
    \includegraphics[width=\textwidth]{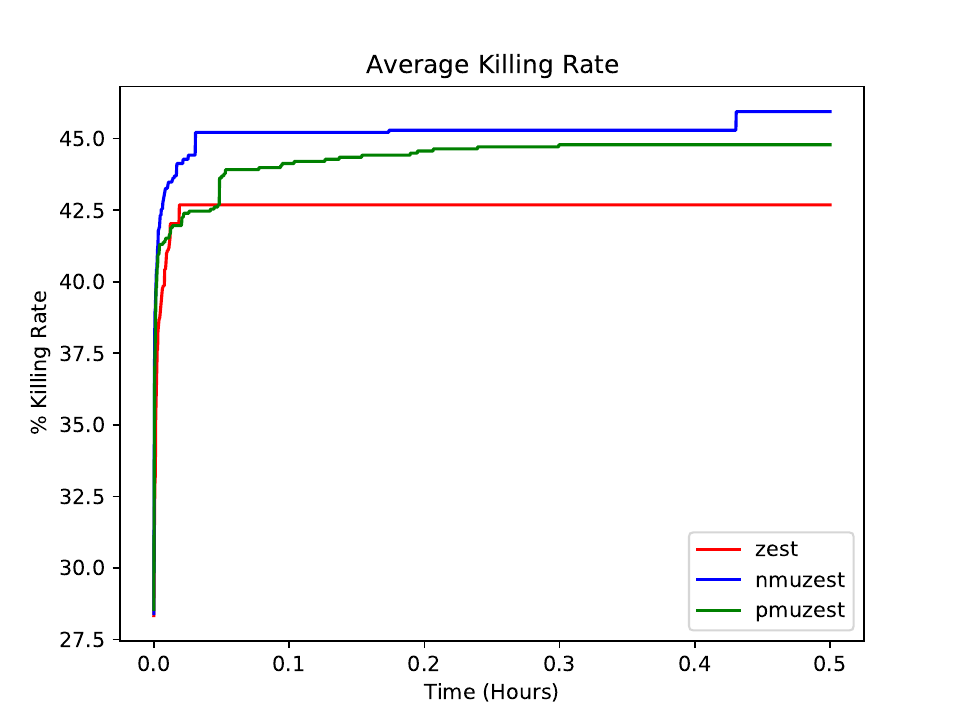}
    \caption{C01-30min}
    \label{subfig:mC01-30min}
\end{subfigure}
\hfill
\begin{subfigure}{0.19\textwidth}
    \includegraphics[width=\textwidth]{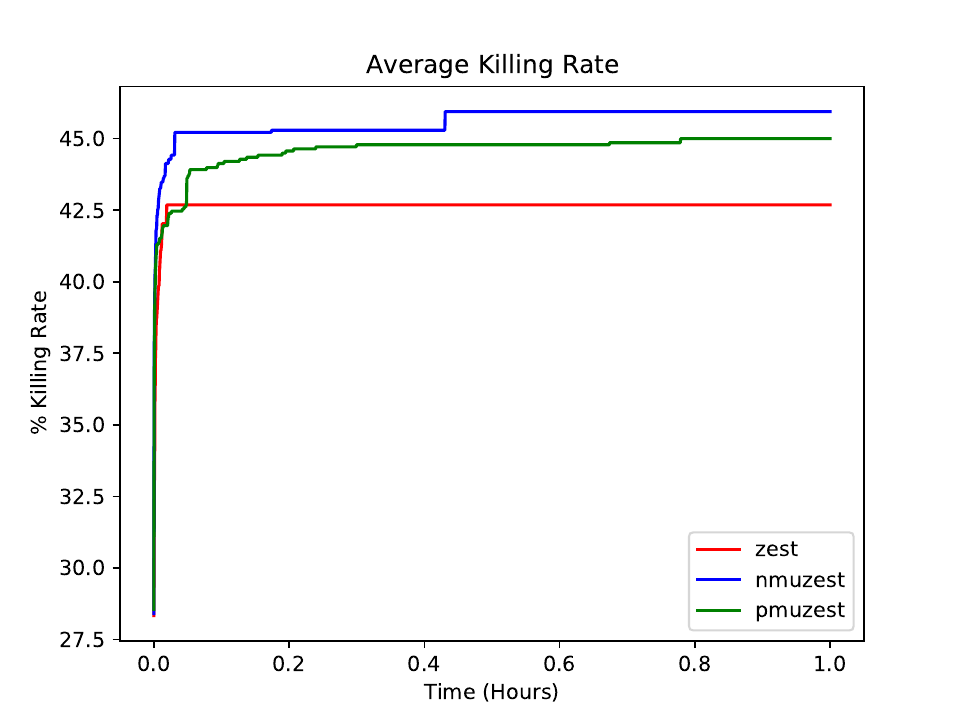}
    \caption{C01-1hour}
    \label{subfig:mC01-1hour}
\end{subfigure}
\hfill
\begin{subfigure}{0.19\textwidth}
    \includegraphics[width=\textwidth]{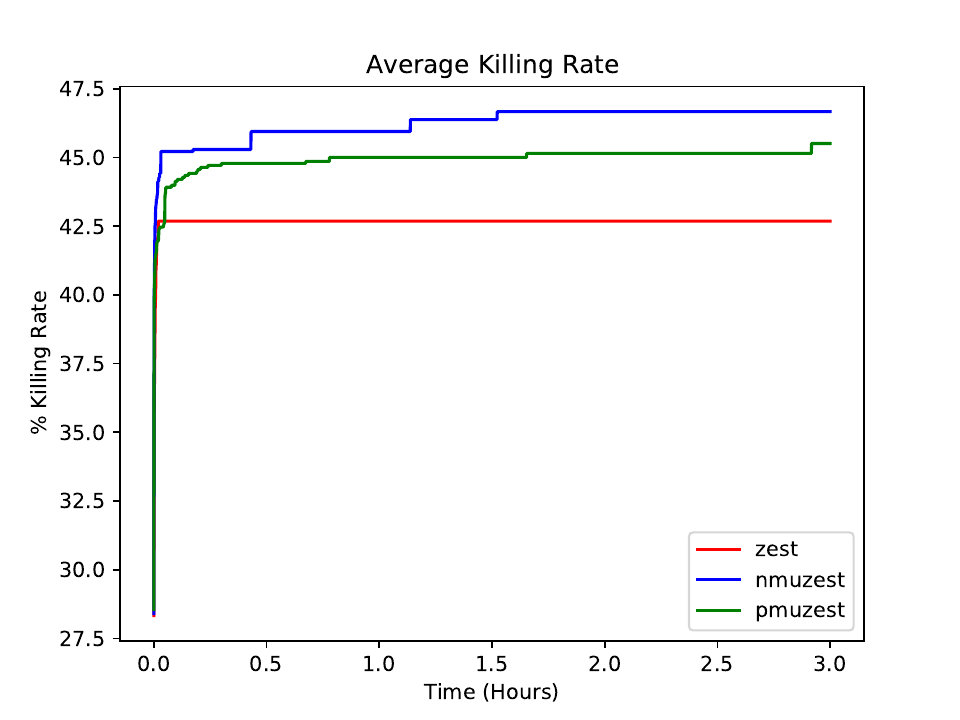}
    \caption{C01-3hour}
    \label{subfig:mC01-3hour}
\end{subfigure}
% -------------------------------------- %
\begin{subfigure}{0.19\textwidth}
    \includegraphics[width=\textwidth]{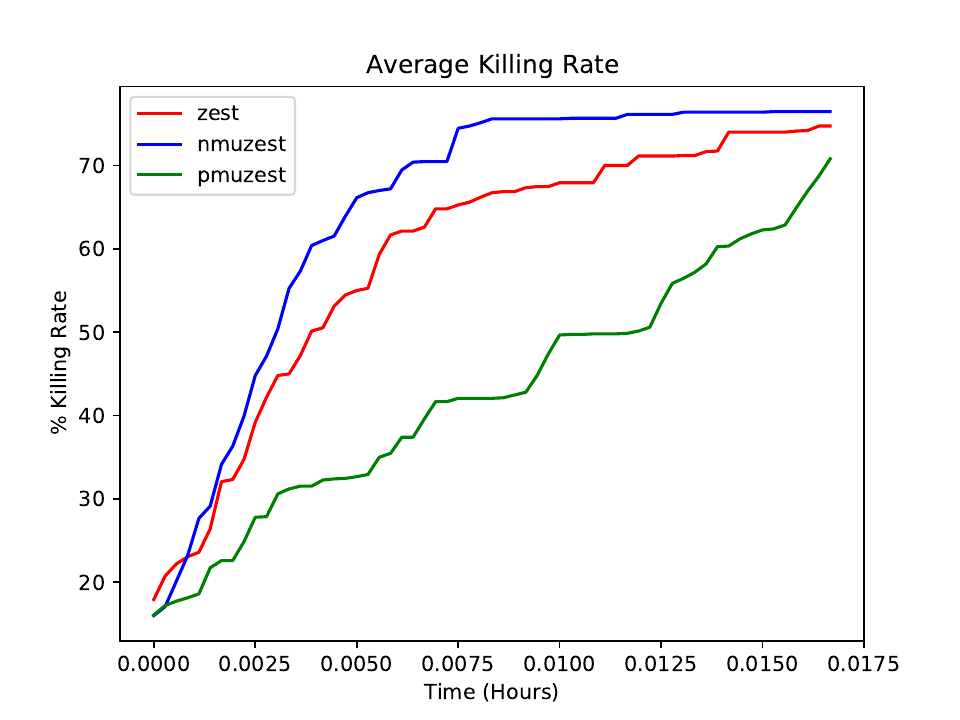}
    \caption{C02-1min}
    \label{subfig:mC02-1min}
\end{subfigure}
\hfill
\begin{subfigure}{0.19\textwidth}
    \includegraphics[width=\textwidth]{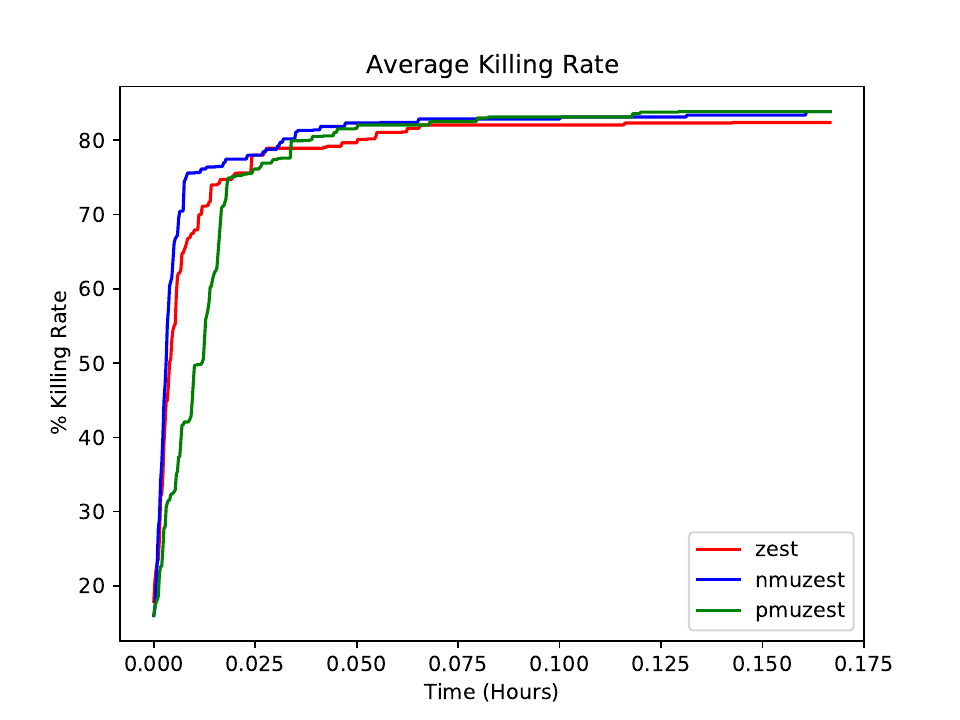}
    \caption{C02-10min}
    \label{subfig:mC02-10min}
\end{subfigure}
\hfill
\begin{subfigure}{0.19\textwidth}
    \includegraphics[width=\textwidth]{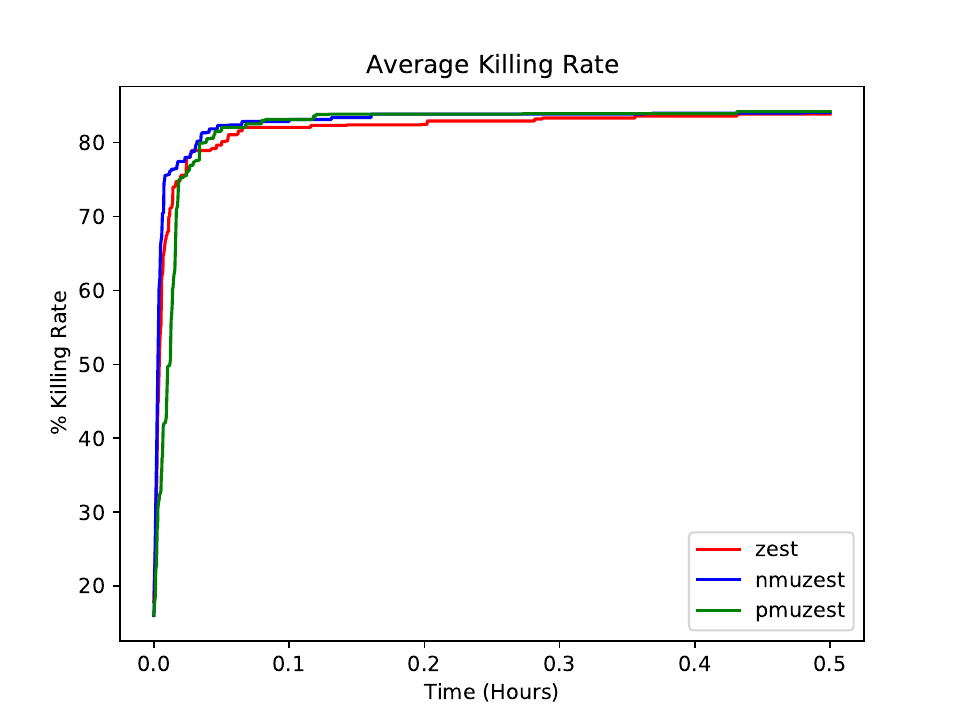}
    \caption{C02-30min}
    \label{subfig:mC02-30min}
\end{subfigure}
\hfill
\begin{subfigure}{0.19\textwidth}
    \includegraphics[width=\textwidth]{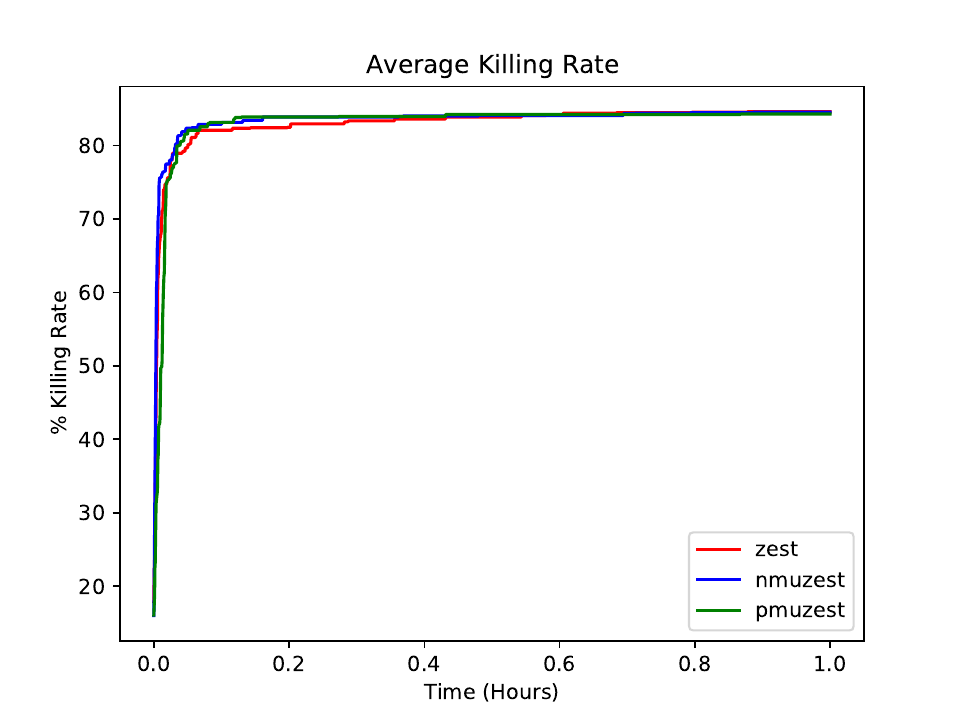}
    \caption{C02-1hour}
    \label{subfig:mC02-1hour}
\end{subfigure}
\hfill
\begin{subfigure}{0.19\textwidth}
    \includegraphics[width=\textwidth]{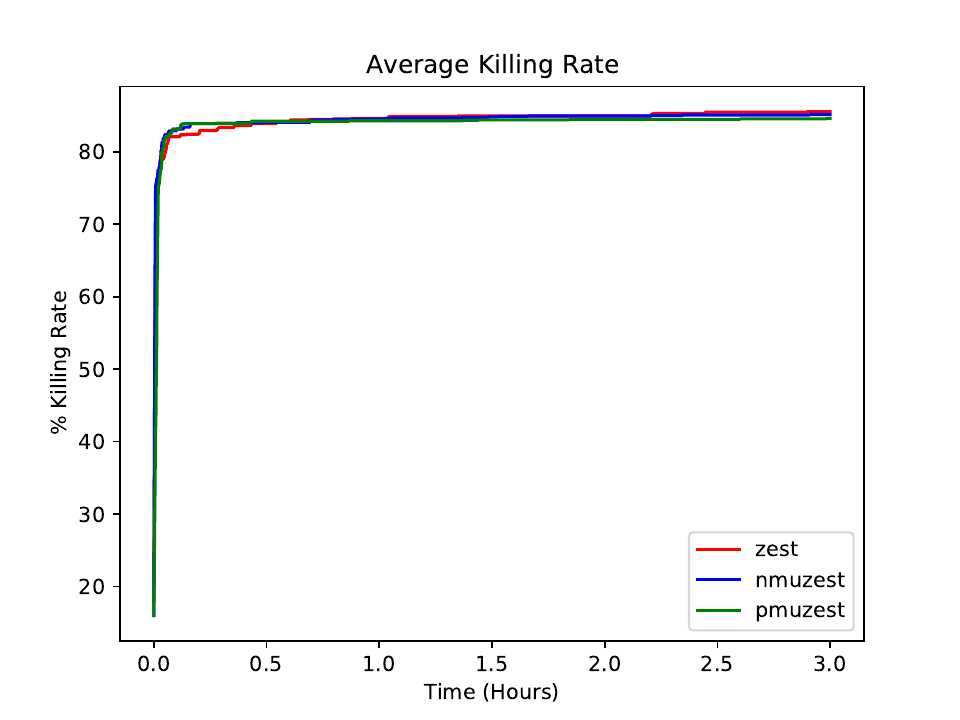}
    \caption{C02-3hour}
    \label{subfig:mC02-3hour}
\end{subfigure}
% -------------------------------------- %
\begin{subfigure}{0.19\textwidth}
    \includegraphics[width=\textwidth]{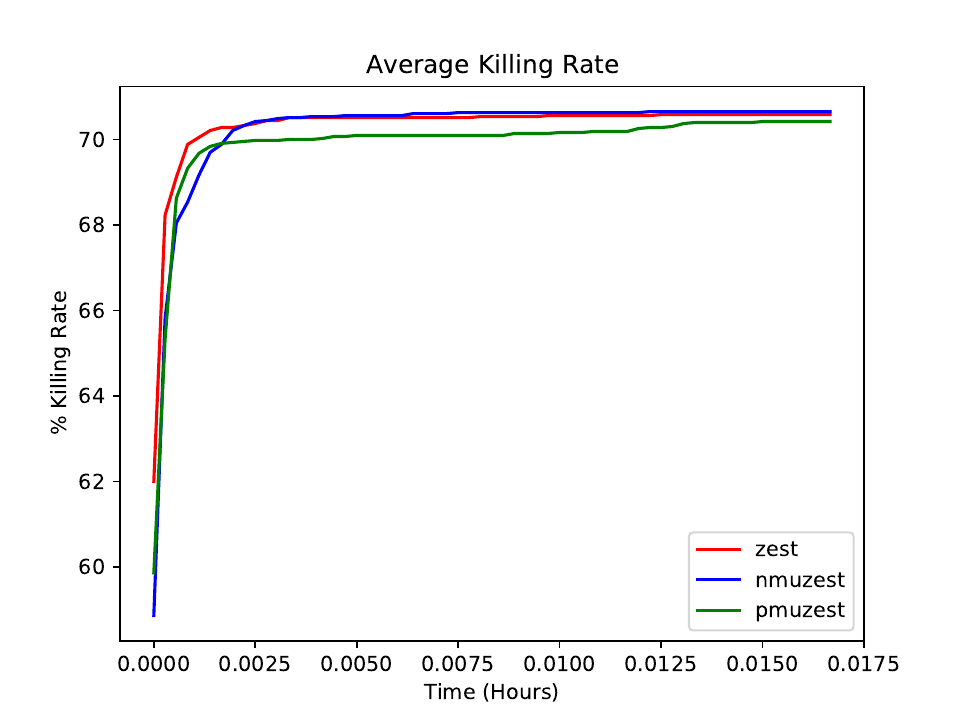}
    \caption{C03-1min}
    \label{subfig:mC03-1min}
\end{subfigure}
\hfill
\begin{subfigure}{0.19\textwidth}
    \includegraphics[width=\textwidth]{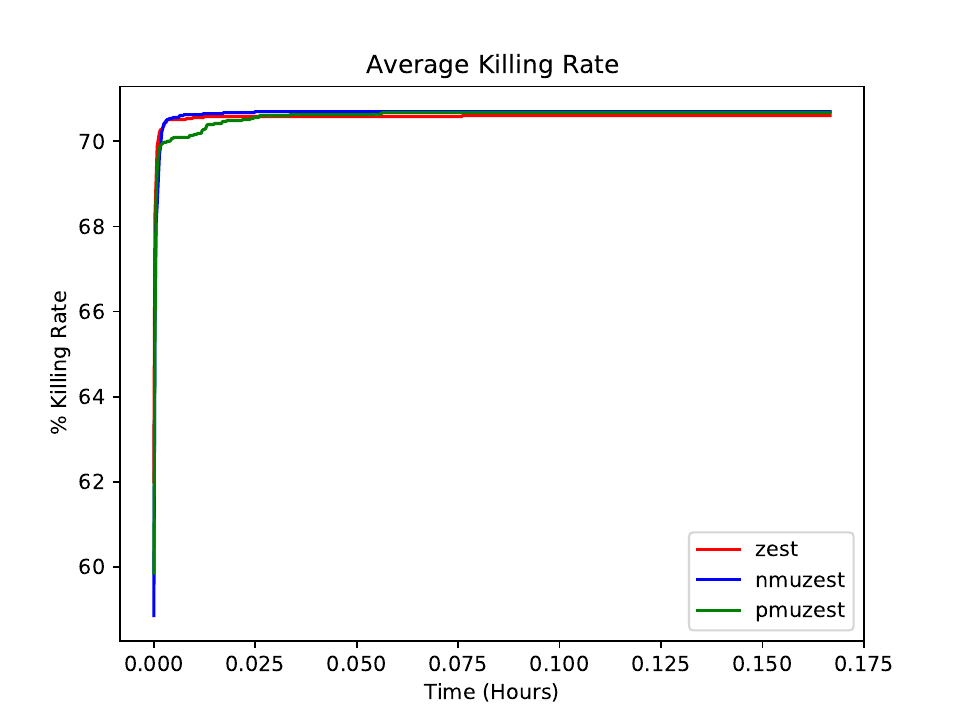}
    \caption{C03-10min}
    \label{subfig:mC03-10min}
\end{subfigure}
\hfill
\begin{subfigure}{0.19\textwidth}
    \includegraphics[width=\textwidth]{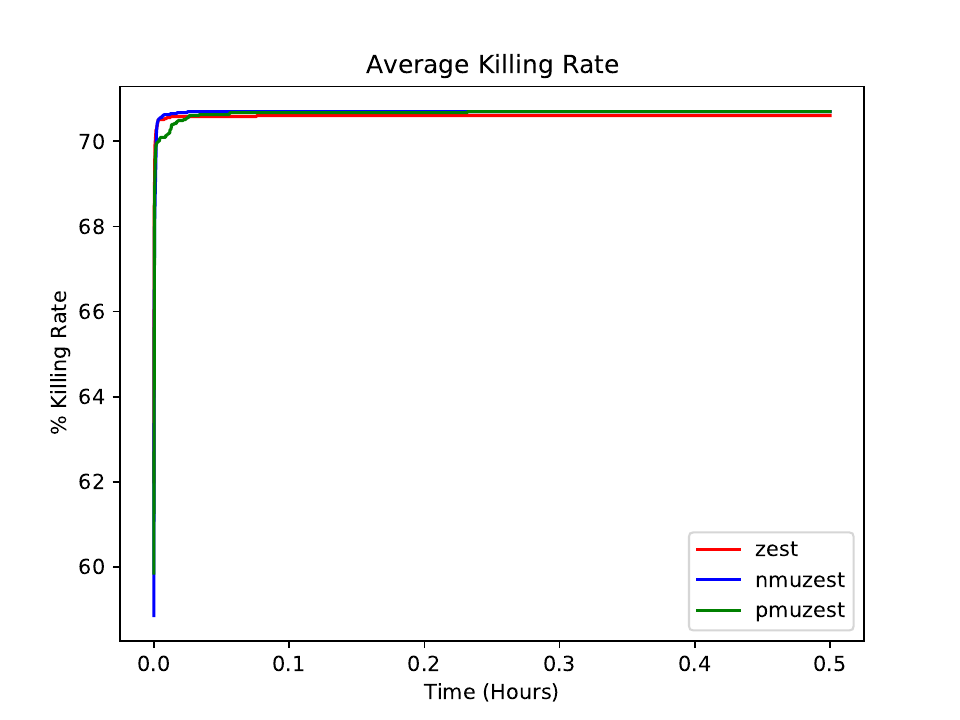}
    \caption{C03-30min}
    \label{subfig:mC03-30min}
\end{subfigure}
\hfill
\begin{subfigure}{0.19\textwidth}
    \includegraphics[width=\textwidth]{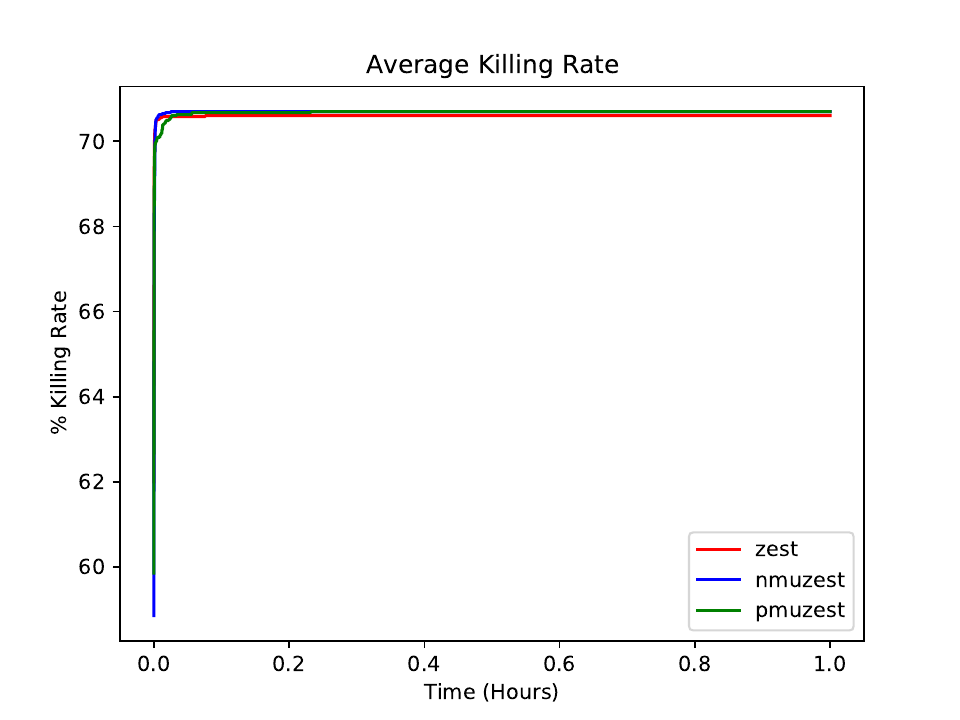}
    \caption{C03-1hour}
    \label{subfig:mC03-1hour}
\end{subfigure}
\hfill
\begin{subfigure}{0.19\textwidth}
    \includegraphics[width=\textwidth]{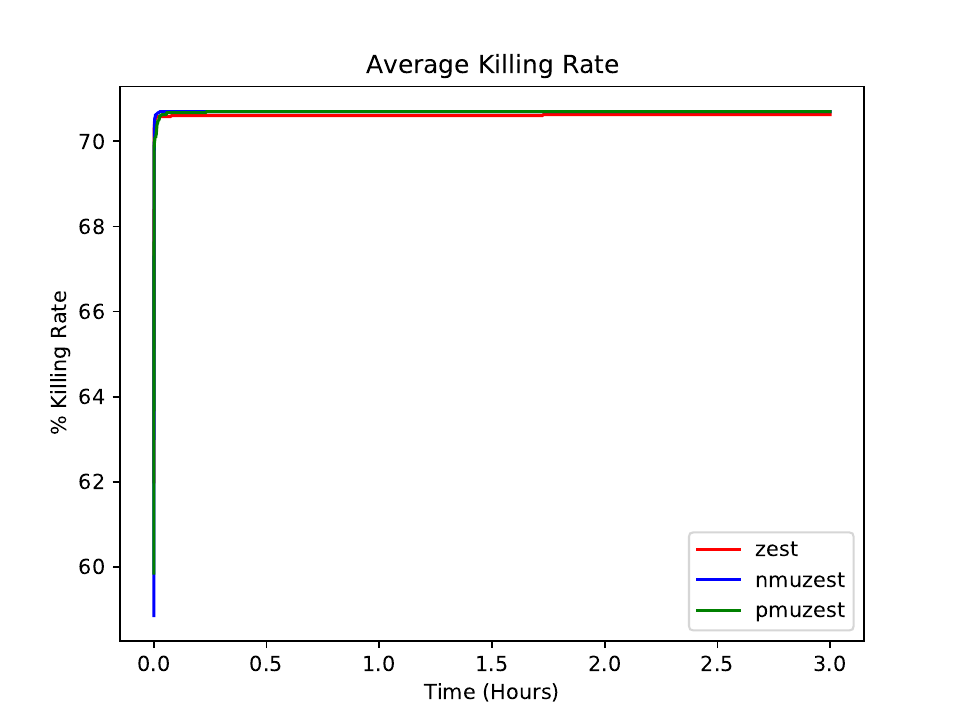}
    \caption{C03-3hour}
    \label{subfig:mC03-1hour}
\end{subfigure}
% -------------------------------------- %
\begin{subfigure}{0.19\textwidth}
    \includegraphics[width=\textwidth]{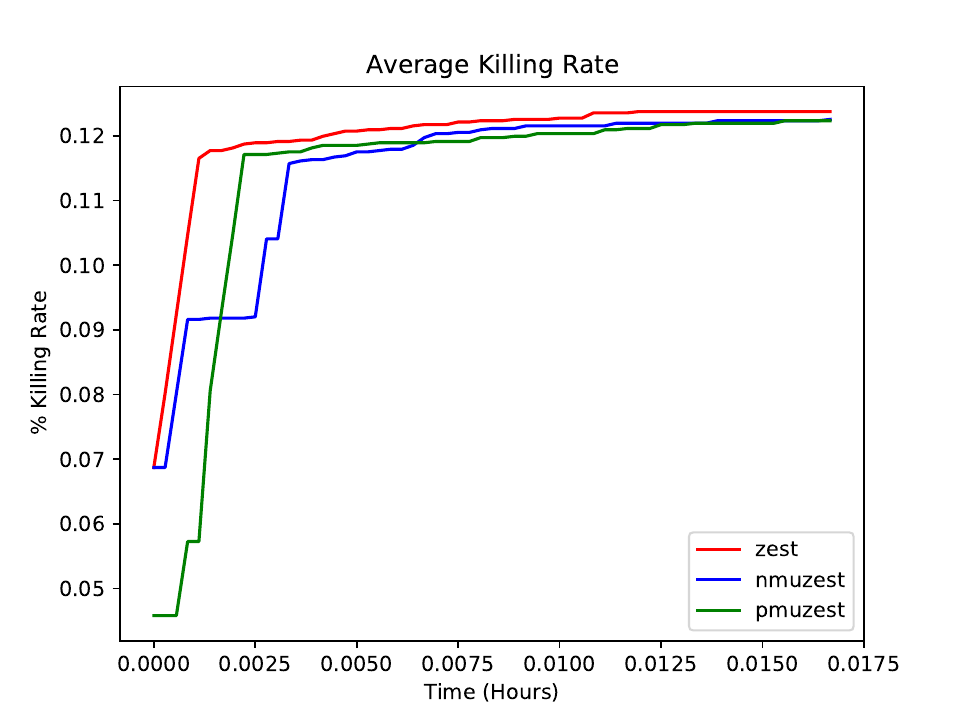}
    \caption{C04-1min}
    \label{subfig:mC04-1min}
\end{subfigure}
\hfill
\begin{subfigure}{0.19\textwidth}
    \includegraphics[width=\textwidth]{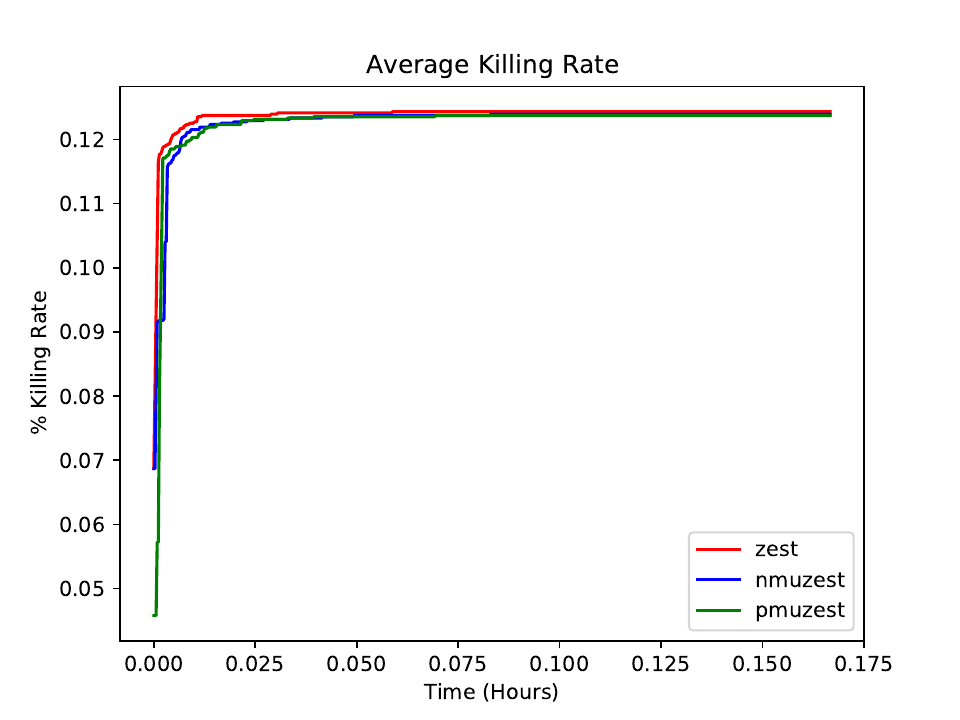}
    \caption{C04-10min}
    \label{subfig:mC04-10min}
\end{subfigure}
\hfill
\begin{subfigure}{0.19\textwidth}
    \includegraphics[width=\textwidth]{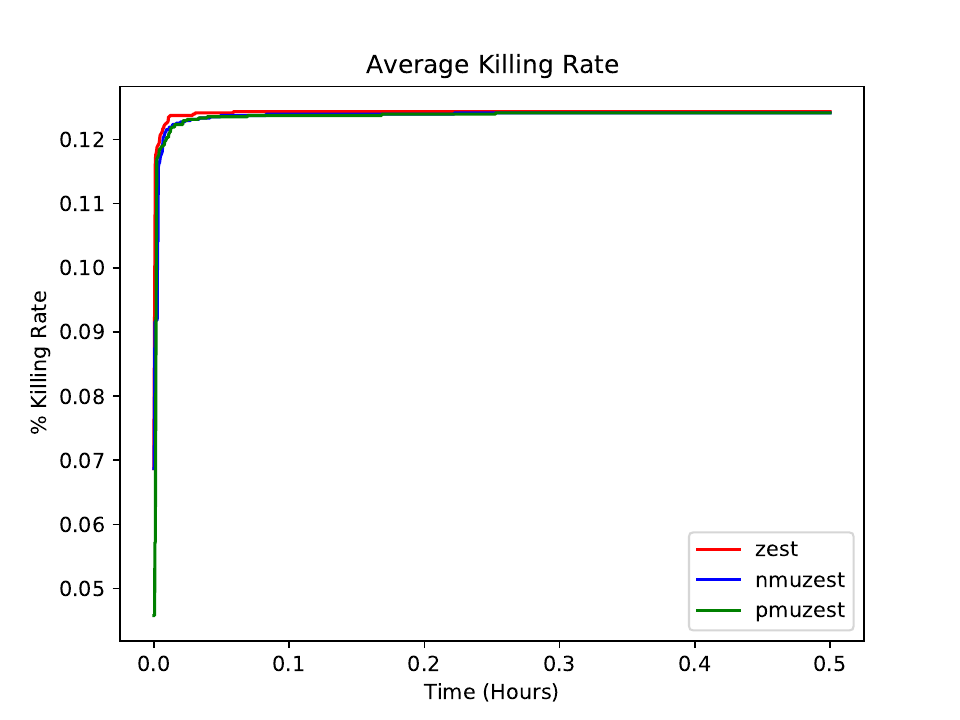}
    \caption{C04-30min}
    \label{subfig:mC04-30min}
\end{subfigure}
\hfill
\begin{subfigure}{0.19\textwidth}
    \includegraphics[width=\textwidth]{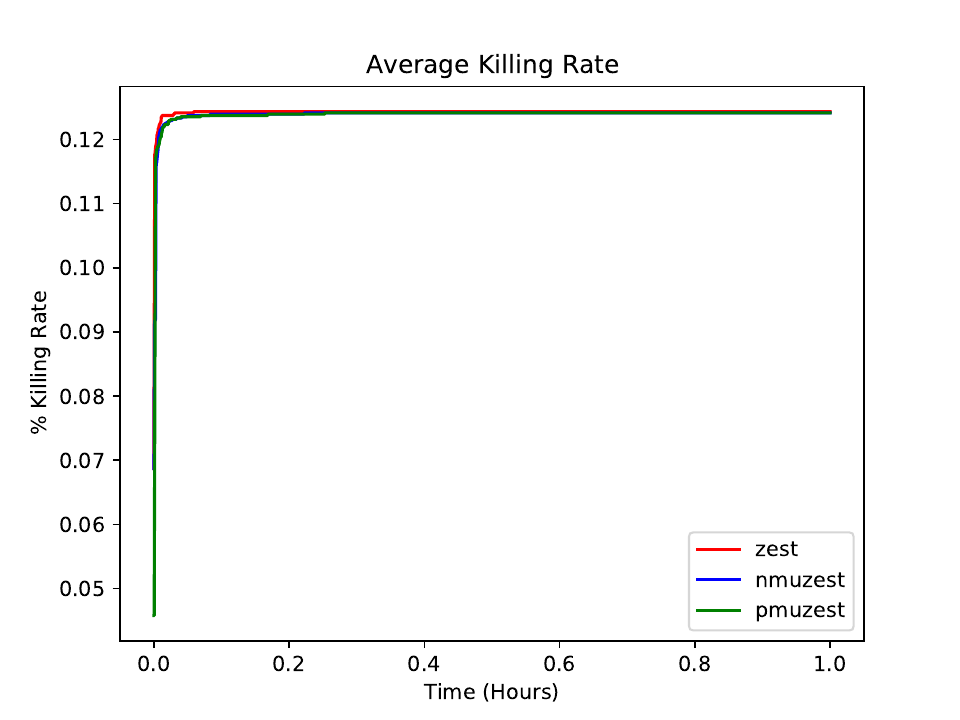}
    \caption{C04-1hour}
    \label{subfig:mC04-1hour}
\end{subfigure}
\hfill
\begin{subfigure}{0.19\textwidth}
    \includegraphics[width=\textwidth]{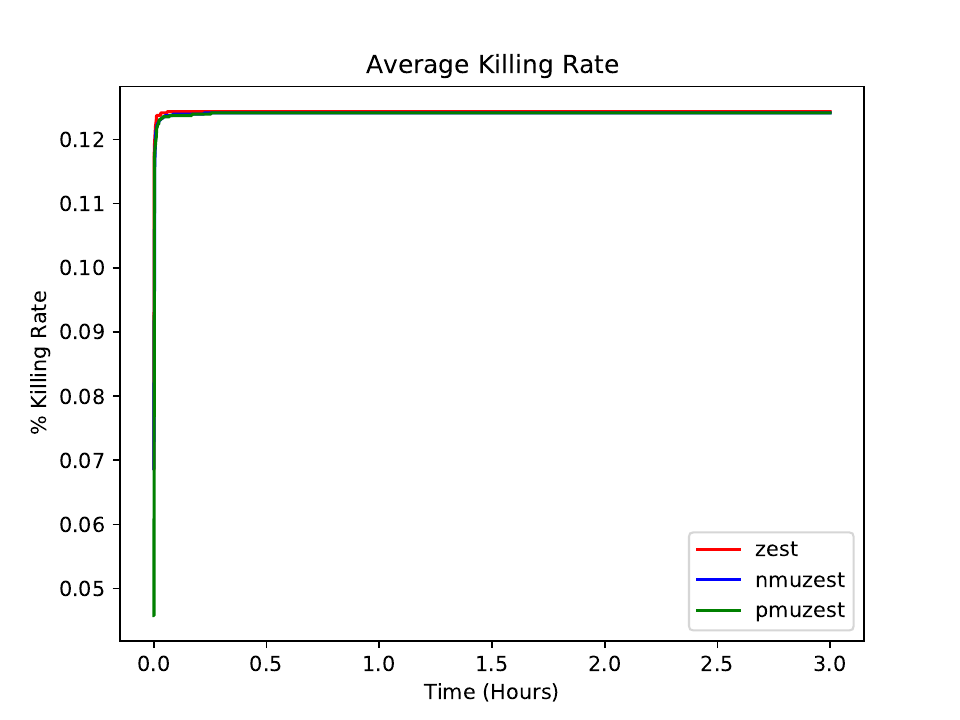}
    \caption{C04-3hour}
    \label{subfig:mC04-3hour}
\end{subfigure}
% -------------------------------------- %
\begin{subfigure}{0.19\textwidth}
    \includegraphics[width=\textwidth]{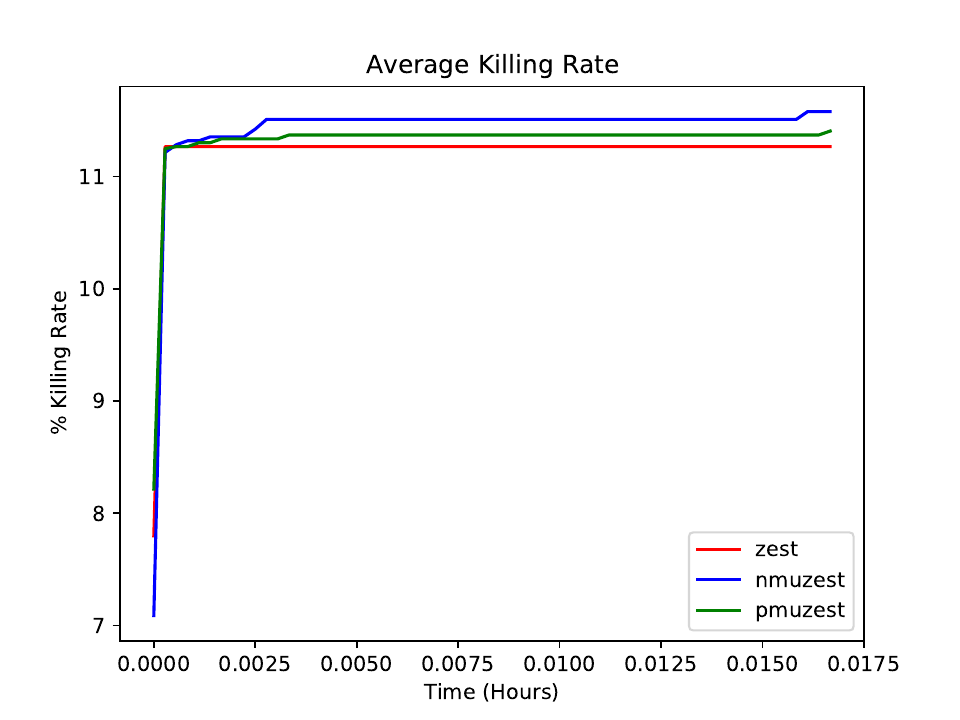}
    \caption{C05-1min}
    \label{subfig:mC05-1min}
\end{subfigure}
\hfill
\begin{subfigure}{0.19\textwidth}
    \includegraphics[width=\textwidth]{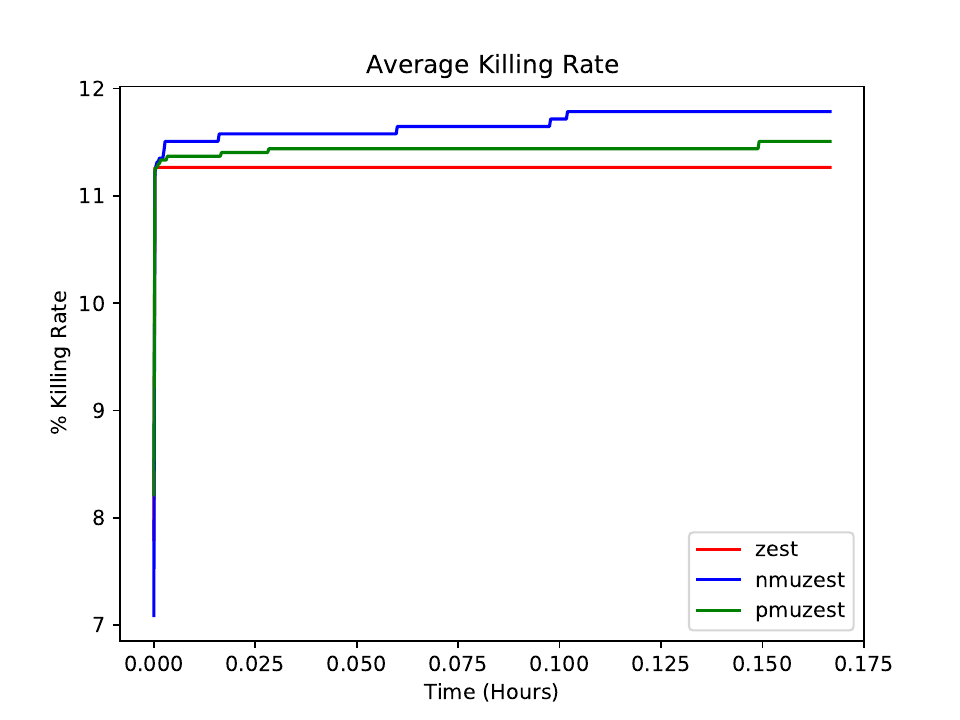}
    \caption{C05-10min}
    \label{subfig:mC05-10min}
\end{subfigure}
\hfill
\begin{subfigure}{0.19\textwidth}
    \includegraphics[width=\textwidth]{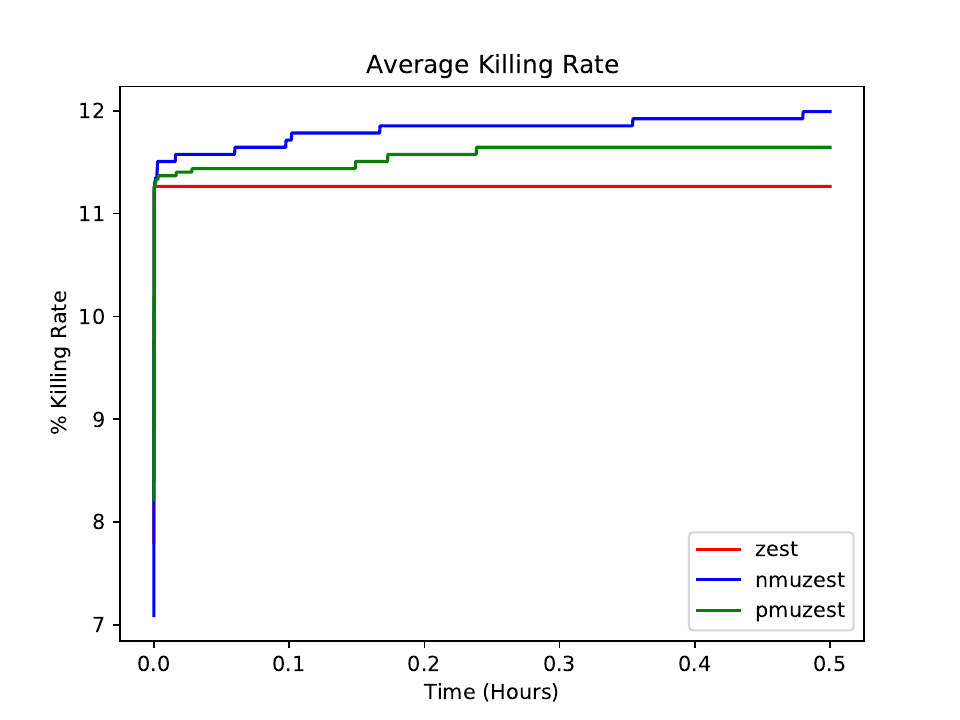}
    \caption{C05-30min}
    \label{subfig:mC05-30min}
\end{subfigure}
\hfill
\begin{subfigure}{0.19\textwidth}
    \includegraphics[width=\textwidth]{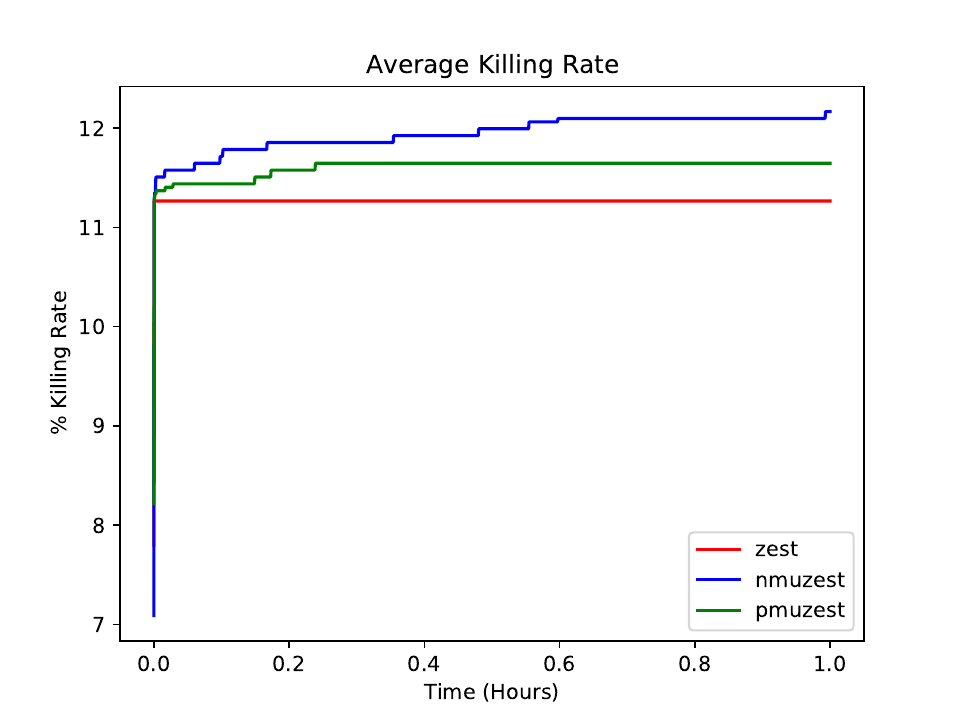}
    \caption{C05-1hour}
    \label{subfig:mC05-1hour}
\end{subfigure}
\hfill
\begin{subfigure}{0.19\textwidth}
    \includegraphics[width=\textwidth]{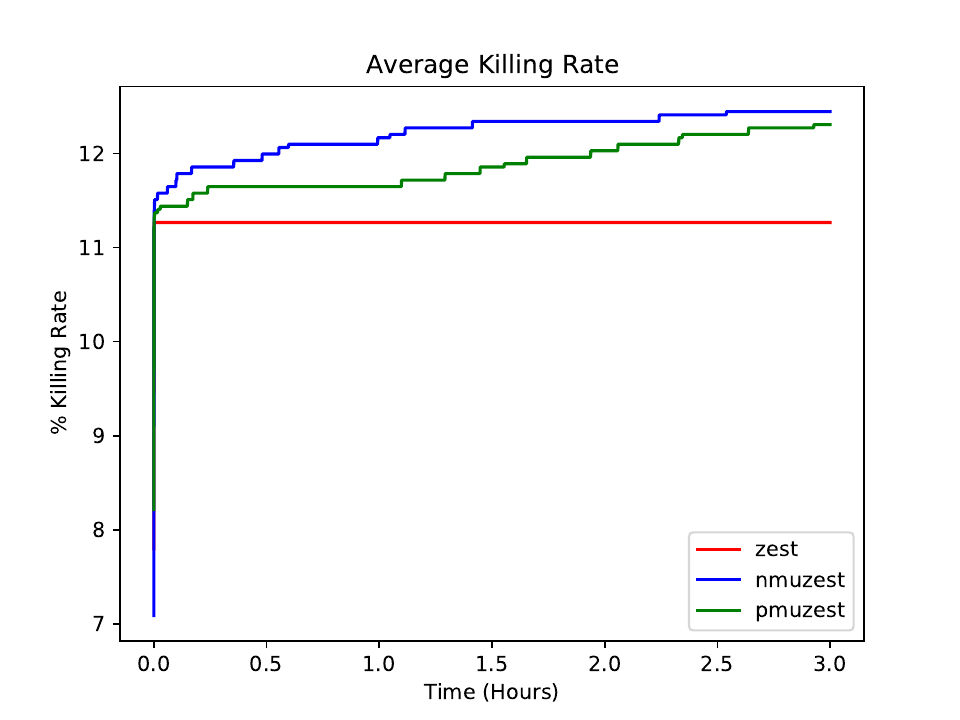}
    \caption{C05-3hour}
    \label{subfig:mC05-3hour}
\end{subfigure}
% -------------------------------------- %
\caption{
Average killing rates for different fuzz campaigns in different time durations.
The x-axis of each figure represents time points and the y-axis represents the achieved mutation killing rates.}
\label{fig:killing_rates}
\end{figure*}

\noindent \textbf{\textit{Analysis.}}
According to the observations above we can see that mutation testing does be able to influence the performance of \zest{} in terms of branch coverage.
The enhancement from opposite directions (\nmuzest{} and \pmuzest{}) presents different results.
Specifically, \nmuzest{} covers more branches compare to \zest{} in some cases.
The negative enhancement with mutation testing directs fuzzing towards covering area far from the mutants that have already been killed, which is consistent to \cite{chan1996proportional}.
However, a positive enhancement with mutation testing (\pmuzest{}) can make fuzzing covering fewer branches (C02 and C04).
This is because \pmuzest{} amplifies the number of children inputs generated from inputs that kill old mutants, which then increases the chance to exploring the branches that has already been covered. Moreover, the maximum branch coverage rates depend on the logic of fuzz cases (which is also know as the distribution of test cases \cite{quickcheck}).
For example, C04 takes a set of double values as inputs to exercise public methods supplies by \texttt{SimpleRegression}.
As a result, C04 is not likely to fuzz branches that belong to types that are not invoked by \texttt{SimpleRegression}.

% \begin{center}
%     \begin{tcolorbox}[
%       colback=gray!10,
%       width=0.47\textwidth,
%       arc=1mm, 
%       auto outer arc,
%       boxrule=0.5pt,
%     ]
    
%     \textbf{\textit{Answers to RQ1.}}
%     Mutation testing can improve the effectiveness of CGF techniques in covering code in a negative manner.
%     % The upper bound of branch  is influenced by the logic of fuzz cases. 
    
%     \end{tcolorbox}
% \end{center}

\finding{1}{
 Mutation testing can improve the effectiveness of CGF techniques in covering code in a negative manner.
 The maximum branches achieved are relevant to the properties of benchmarks.
}

\subsection{RQ2: Effectiveness in killing Mutants}

Figure \ref{fig:killing_rates} shows average killing rates achieved during fuzz campaigns.
Like analysis for branch coverage, Figure \ref{fig:killing_rates} also illustrates curves in different time durations.
The y-axis of the sub-figures represents the average killing rates.
From Figure \ref{fig:killing_rates} we can also get observations from the following perspectives:
\begin{itemize}[leftmargin=*, topsep=3pt]
    \item \textit{\textbf{Compare among durations.}} Like branch coverage, the increases of mutation killing rates also happen at the beginning of each fuzz campaigns.
    However, killing rates curves go flat at much latter time points compares to branch coverage.
    In C01 and C05, curves for \nmuzest{} and \pmuzest{} still rise after fuzzing for more than 1 hour. 
    
    \item \textit{\textbf{Compare among techniques.}}
    Both the enhanced techniques (\nmuzest{} and \pmuzest{}) are better than the baseline technique \zest{}, and \nmuzest{} is better than \pmuzest{}.
    In C01 and C05, the killing rates of the enhanced techniques rise faster than \zest{} and finally come to a higher killing rates.
    In C02, \zest{} is worse than \nmuzest{} but better than \pmuzest{} during the first 10 minutes. 
    Three curves finally merge together after fuzzing for 30 minutes.
    In C03 and C04, although the killing rate of \zest{} rises fastest in the first minute, it is immediately caught up by the enhanced techniques in the next few minutes. 
    
    \item \textit{\textbf{Compare among cases.}}
    We get observations similar to Figure \ref{fig:branch_covs} for killing rates.
    The increasing tendencies and the maximum kill rates also varies for different fuzz cases. 
    The killing rate curves for C02$\sim$C04 rise drastically at the first minute of fuzzing campaigns and rapidly become flat in the next few minutes, 
    whereas for C01 and C05 the curves keep increasing for the while 3 hours.
    In terms of maximum kill rates,
    in C02 and C03 all the three techniques can achieve peek values that are larger than 70\%.
    C04 achieves the smallest maximum kill rates around 0.12\%. 
\end{itemize}

\textit{\textbf{Analysis.}}
Compares to \zest{}, the enhanced techniques kill more mutants with the guidance of mutation testing.
However, although both \nmuzest{} and \pmuzest{} achieve higher kill rates than \zest{}, the type of the mutants which are additionally killed should be different.
% We make detailed analysis for killed mutants in Section \ref{subsec:killed_mutants}.
Moreover, the object for fuzzing as well as the logic of fuzz cases are also important factors for killing mutants. 

% \begin{center}
%     \begin{tcolorbox}[
%       colback=gray!10,
%       width=0.47\textwidth,
%       arc=1mm, 
%       auto outer arc,
%       boxrule=0.5pt,
%     ]
    
%     \textbf{\textit{Answers to RQ2.}}
%     Mutation testing can improve the effectiveness of CGF techniques in killing mutants in both negative and positive manners.
%     % The upper bound of killing mutants is relevant to the property of objects of fuzzing and the property of fuzz cases. 
    
%     \end{tcolorbox}
% \end{center}

\finding{2}{
Mutation testing can improve the effectiveness of CGF techniques in killing mutants with both negative and positive manners.
The maximum numbers of mutants killed are relevant to the properties of benchmarks. 
}

\section{Related Work}
In this paper, we enhances CGF via identifying bug-revealing inputs and amplifying their effect by mutation testing.
Our work is related to \textit{coverage guided fuzzing}, \textit{mutation testing} and \textit{differential testing}.

% \noindent
\textit{Coverage Guided Fuzzing.}
% \subsubsection*{Coverage Guided Fuzzing}
Fuzzing has been a hot research area in the past few decades \cite{fuzzSurvey}.
It was introduced in the early 1990s \cite{DBLP:journals/cacm/MillerFS90}.
At first, fuzzing was dedicated for testing a PUT fully randomly with given inputs, which called seeds.
To make fuzzing more systematic, CGF guides fuzzing with code coverage attained with lightweight instrumentation.
AFL \cite{afl} employs a novel type of compile-time instrumentation and genetic algorithms to discover interesting test inputs. 
% It guides fuzzing towards triggering new internal states in the targeted binary with functional coverage attained by instrumentation.
Based on AFL, AFLFast \cite{aflfast} proposes strategies to systematically bias fuzzer towards exercising low-frequency paths.
Libfuzzer \cite{libfuzz} is an in-process, coverage-guided, evolutionary fuzzer which linked with the library under test, and feeds fuzzed inputs to the library via a specific fuzzing entrypoint.
It tracks reached code areas and generates inputs by mutating the corpus of seeds in order to maximize code coverage.
Both AFL \cite{afl} and Libfuzzer \cite{libfuzz} are typical coverage guided fuzzers which are widely used in literature \cite{muzz, zest, aflfast, memlock, qsym}.
% \cite{muzz, zest, aflfast, memlock, qsym,  DBLP:conf/desec/ChaoLCTH18, DBLP:conf/secdev/Serebryany16}.

% Enhancement to CGF: JQF -> Zest, PerfFuzz, BigFuzz, MoFuzz
Some researches introduce extra information coverage to facilitate fuzzing on finding specific types of bugs.
MUZZ \cite{muzz} provides three types of instruments to discover and reveal multi-thread vulnerabilities of the PUT during fuzzing.
MemLock \cite{memlock} tracks memory usage during fuzzing in order to trigger uncontrolled memory usage bugs. PerfFuzz \cite{perffuzz} endeavors to generate pathological inputs which exercise hot spots of the PUT or with a higher total execution path length.
To trigger vulnerabilities at semantic stage, Zest \cite{zest} proposes an approach with property-based testing \cite{quickcheck} which build and mutate inputs in a semantic-valid way in order to detect vulnerabilities at semantic stage.
This paper aims to enhance CGF to discover more bug-revealing inputs enhanced by mutation testing, which is distinct from the researches mentioned above. 

% \noindent
\textit{Mutation Testing.}
% Researches use mutation testing
% \subsubsection*{Mutation Testing}
Mutation testing measures the adequacy of testing with \textit{mutation score}, or \textit{mutation coverage}.
In this regard, many researches use mutation testing as a manner of feedback or guidance. 
For example, Mike and Yves \cite{DBLP:journals/stvr/PapadakisT15} build a mutation-based fault local technique with a test suite constructed from mutation testing.
Gordon and Andreas \cite{DBLP:conf/issta/FraserZ10} use mutation testing to guide oracle constructions.
Some test optimization techniques also benefit from mutation testing, such as test suite reduction \cite{DBLP:conf/sigsoft/ShiGGZM14} and test case prioritization \cite{DBLP:conf/issre/LouH015}.
Unlike these techniques, the goal of our approach is to use mutation testing to guide fuzzing towards finding inputs that are capable of finding bugs.

% \noindent
\textit{Differential Testing.}
% \subsubsection*{Differential Testing}
Differential testing \cite{98difftest} detects bugs via comparing comparable systems.
It amends the absence of oracles by using the outputs of different systems to validate the behaviors of themselves interactively.
Differential testing is widely adopted in situations that oracles are hard to obtain, such as 
\textit{compiler testing} 
\cite{DBLP:conf/pldi/YeTTHFSBW021, DBLP:conf/icse/ParkAYKR21}, 
% \cite{DBLP:conf/pldi/YeTTHFSBW021, DBLP:conf/icse/ParkAYKR21, DBLP:conf/icse/ChenSS19}, 
\textit{DNN testing}\cite{DBLP:conf/sigsoft/AsyrofiY021, DBLP:conf/sigsoft/MaYK21}
% \cite{DBLP:conf/sigsoft/AsyrofiY021, DBLP:conf/sigsoft/MaYK21, DBLP:conf/sigsoft/GuoJZCS18}, 
and \textit{regression testing} \cite{ DBLP:conf/icse/GulzarZH19, DBLP:conf/issta/GodefroidLP20}.
% \cite{DBLP:conf/sigsoft/EvansS07, DBLP:conf/kbse/TanejaX08, DBLP:conf/icse/GulzarZH19, DBLP:conf/issta/GodefroidLP20}.
In this paper, we utilize differential testing to enable the detection of mutants.
We propose a configurable differential outputs checking frameworks. It checks the consistency of outputs during fuzzing.
Mutants of which the outputs are inconsistent to those of the PUT is detected and will be considered as killed.

\section{Conclusion}
In this paper, we incorporate mutation testing with fuzzing in order to guide fuzzing towards detecting bugs.
We conduct a well-designed experiment on \nBenchmark{} benchmarks with 3 techniques (2 modified techniques and 1 baseline) to evaluate the proposed approach.
The experimental results show that mutation testing can make progress in terms of both code coverage and bug detection.

Our approach is general and can be extended to other CGF techniques.
A larger scale experiments are worth conducting on other real world benchmarks as well as coverage guided fuzzers.
Meanwhile, it is also important to investigate how different types of mutants can influence our approach.
We leave these as future works.

\section*{Acknowledgments}
The authors would like to thank the anonymous reviewers for insightful comments.
This research is partially supported by the National Natural Science Foundation of China (No. 61932012, 62141215).

% \vspace{-2mm}
\bibliographystyle{ACM-Reference-Format}
\bibliography{bibfile}


\begin{thebibliography}{33}

%%% the Web address.

\ifx \showCODEN    \undefined \def \showCODEN     #1{\unskip}     \fi
\ifx \showDOI      \undefined \def \showDOI       #1{#1}\fi
\ifx \showISBNx    \undefined \def \showISBNx     #1{\unskip}     \fi
\ifx \showISBNxiii \undefined \def \showISBNxiii  #1{\unskip}     \fi
\ifx \showISSN     \undefined \def \showISSN      #1{\unskip}     \fi
\ifx \showLCCN     \undefined \def \showLCCN      #1{\unskip}     \fi
\ifx \shownote     \undefined \def \shownote      #1{#1}          \fi
\ifx \showarticletitle \undefined \def \showarticletitle #1{#1}   \fi
\ifx \showURL      \undefined \def \showURL       {\relax}        \fi
% invisible to TeX
\providecommand\bibfield[2]{#2}
\providecommand\bibinfo[2]{#2}
\providecommand\natexlab[1]{#1}
\providecommand\showeprint[2][]{arXiv:#2}

\bibitem[\protect\citeauthoryear{Asyrofi, Yang, and Lo}{Asyrofi
  et~al\mbox{.}}{2021}]%
        {DBLP:conf/sigsoft/AsyrofiY021}
\bibfield{author}{\bibinfo{person}{Muhammad~Hilmi Asyrofi},
  \bibinfo{person}{Zhou Yang}, {and} \bibinfo{person}{David Lo}.}
  \bibinfo{year}{2021}\natexlab{}.
\newblock \showarticletitle{Crossasr++: A modular differential testing
  framework for automatic speech recognition}. In
  \bibinfo{booktitle}{\emph{Proceedings of the 29th ACM Joint Meeting on
  European Software Engineering Conference and Symposium on the Foundations of
  Software Engineering}}. \bibinfo{pages}{1575--1579}.
\newblock


\bibitem[\protect\citeauthoryear{B{\"{o}}hme, Pham, and
  Roychoudhury}{B{\"{o}}hme et~al\mbox{.}}{2019}]%
        {aflfast}
\bibfield{author}{\bibinfo{person}{Marcel B{\"{o}}hme},
  \bibinfo{person}{Van{-}Thuan Pham}, {and} \bibinfo{person}{Abhik
  Roychoudhury}.} \bibinfo{year}{2019}\natexlab{}.
\newblock \showarticletitle{Coverage-Based Greybox Fuzzing as Markov Chain}.
\newblock \bibinfo{journal}{\emph{{IEEE} Trans. Software Eng.}}
  \bibinfo{volume}{45}, \bibinfo{number}{5} (\bibinfo{year}{2019}),
  \bibinfo{pages}{489--506}.
\newblock


\bibitem[\protect\citeauthoryear{Chan, Chen, Mak, and Yu}{Chan
  et~al\mbox{.}}{1996}]%
        {chan1996proportional}
\bibfield{author}{\bibinfo{person}{FT Chan}, \bibinfo{person}{Tsong~Yueh Chen},
  \bibinfo{person}{IK Mak}, {and} \bibinfo{person}{Yuen-Tak Yu}.}
  \bibinfo{year}{1996}\natexlab{}.
\newblock \showarticletitle{Proportional sampling strategy: guidelines for
  software testing practitioners}.
\newblock \bibinfo{journal}{\emph{Information and Software Technology}}
  \bibinfo{volume}{38}, \bibinfo{number}{12} (\bibinfo{year}{1996}),
  \bibinfo{pages}{775--782}.
\newblock


\bibitem[\protect\citeauthoryear{Chen, Guo, Xue, Sui, Zhang, Li, Wang, and
  Liu}{Chen et~al\mbox{.}}{2020}]%
        {muzz}
\bibfield{author}{\bibinfo{person}{Hongxu Chen}, \bibinfo{person}{Shengjian
  Guo}, \bibinfo{person}{Yinxing Xue}, \bibinfo{person}{Yulei Sui},
  \bibinfo{person}{Cen Zhang}, \bibinfo{person}{Yuekang Li},
  \bibinfo{person}{Haijun Wang}, {and} \bibinfo{person}{Yang Liu}.}
  \bibinfo{year}{2020}\natexlab{}.
\newblock \showarticletitle{{MUZZ:} Thread-aware Grey-box Fuzzing for Effective
  Bug Hunting in Multithreaded Programs}. In \bibinfo{booktitle}{\emph{29th
  {USENIX} Security Symposium, {USENIX} Security 2020, August 12-14, 2020}}.
  \bibinfo{publisher}{{USENIX} Association}, \bibinfo{pages}{2325--2342}.
\newblock


\bibitem[\protect\citeauthoryear{Claessen and Hughes}{Claessen and
  Hughes}{2000}]%
        {quickcheck}
\bibfield{author}{\bibinfo{person}{Koen Claessen} {and} \bibinfo{person}{John
  Hughes}.} \bibinfo{year}{2000}\natexlab{}.
\newblock \showarticletitle{QuickCheck: a lightweight tool for random testing
  of Haskell programs}. In \bibinfo{booktitle}{\emph{Proceedings of the Fifth
  {ACM} {SIGPLAN} International Conference on Functional Programming {(ICFP}
  '00), Montreal, Canada, September 18-21, 2000}}. \bibinfo{publisher}{{ACM}},
  \bibinfo{pages}{268--279}.
\newblock


\bibitem[\protect\citeauthoryear{Coles, Laurent, Henard, Papadakis, and
  Ventresque}{Coles et~al\mbox{.}}{2016}]%
        {pitest}
\bibfield{author}{\bibinfo{person}{Henry Coles}, \bibinfo{person}{Thomas
  Laurent}, \bibinfo{person}{Christopher Henard}, \bibinfo{person}{Mike
  Papadakis}, {and} \bibinfo{person}{Anthony Ventresque}.}
  \bibinfo{year}{2016}\natexlab{}.
\newblock \showarticletitle{{PIT:} a practical mutation testing tool for Java
  (demo)}. In \bibinfo{booktitle}{\emph{Proceedings of the 25th International
  Symposium on Software Testing and Analysis, {ISSTA} 2016, Saarbr{\"{u}}cken,
  Germany, July 18-20, 2016}}. \bibinfo{publisher}{{ACM}},
  \bibinfo{pages}{449--452}.
\newblock


\bibitem[\protect\citeauthoryear{Fraser and Zeller}{Fraser and Zeller}{2010}]%
        {DBLP:conf/issta/FraserZ10}
\bibfield{author}{\bibinfo{person}{Gordon Fraser} {and}
  \bibinfo{person}{Andreas Zeller}.} \bibinfo{year}{2010}\natexlab{}.
\newblock \showarticletitle{Mutation-driven generation of unit tests and
  oracles}. In \bibinfo{booktitle}{\emph{Proceedings of the Nineteenth
  International Symposium on Software Testing and Analysis, {ISSTA} 2010,
  Trento, Italy, July 12-16, 2010}}. \bibinfo{publisher}{{ACM}},
  \bibinfo{pages}{147--158}.
\newblock


\bibitem[\protect\citeauthoryear{Ganesh, Leek, and Rinard}{Ganesh
  et~al\mbox{.}}{2009}]%
        {DBLP:conf/icse/GaneshLR09}
\bibfield{author}{\bibinfo{person}{Vijay Ganesh}, \bibinfo{person}{Tim Leek},
  {and} \bibinfo{person}{Martin~C. Rinard}.} \bibinfo{year}{2009}\natexlab{}.
\newblock \showarticletitle{Taint-based directed whitebox fuzzing}. In
  \bibinfo{booktitle}{\emph{31st International Conference on Software
  Engineering, {ICSE} 2009, May 16-24, 2009, Vancouver, Canada, Proceedings}}.
  \bibinfo{publisher}{{IEEE}}, \bibinfo{pages}{474--484}.
\newblock


\bibitem[\protect\citeauthoryear{github}{github}{arch}]%
        {github}
\bibfield{author}{\bibinfo{person}{github}.} \bibinfo{year}{visited at 2022
  March}\natexlab{}.
\newblock \bibinfo{title}{Github}.
\newblock \bibinfo{howpublished}{\url{https://github.com/}}.
\newblock


\bibitem[\protect\citeauthoryear{Gligoric, Groce, Zhang, Sharma, Alipour, and
  Marinov}{Gligoric et~al\mbox{.}}{2015}]%
        {DBLP:journals/tosem/GligoricGZSAM15}
\bibfield{author}{\bibinfo{person}{Milos Gligoric}, \bibinfo{person}{Alex
  Groce}, \bibinfo{person}{Chaoqiang Zhang}, \bibinfo{person}{Rohan Sharma},
  \bibinfo{person}{Mohammad~Amin Alipour}, {and} \bibinfo{person}{Darko
  Marinov}.} \bibinfo{year}{2015}\natexlab{}.
\newblock \showarticletitle{Guidelines for Coverage-Based Comparisons of
  Non-Adequate Test Suites}.
\newblock \bibinfo{journal}{\emph{{ACM} Trans. Softw. Eng. Methodol.}}
  \bibinfo{volume}{24}, \bibinfo{number}{4} (\bibinfo{year}{2015}),
  \bibinfo{pages}{22:1--22:33}.
\newblock


\bibitem[\protect\citeauthoryear{Godefroid, Lehmann, and Polishchuk}{Godefroid
  et~al\mbox{.}}{2020}]%
        {DBLP:conf/issta/GodefroidLP20}
\bibfield{author}{\bibinfo{person}{Patrice Godefroid}, \bibinfo{person}{Daniel
  Lehmann}, {and} \bibinfo{person}{Marina Polishchuk}.}
  \bibinfo{year}{2020}\natexlab{}.
\newblock \showarticletitle{Differential regression testing for {REST} APIs}.
  In \bibinfo{booktitle}{\emph{{ISSTA} '20: 29th {ACM} {SIGSOFT} International
  Symposium on Software Testing and Analysis, Virtual Event, USA, July 18-22,
  2020}}. \bibinfo{publisher}{{ACM}}, \bibinfo{pages}{312--323}.
\newblock


\bibitem[\protect\citeauthoryear{Godefroid, Levin, and Molnar}{Godefroid
  et~al\mbox{.}}{2012}]%
        {SAGE}
\bibfield{author}{\bibinfo{person}{Patrice Godefroid},
  \bibinfo{person}{Michael~Y. Levin}, {and} \bibinfo{person}{David~A. Molnar}.}
  \bibinfo{year}{2012}\natexlab{}.
\newblock \showarticletitle{{SAGE:} whitebox fuzzing for security testing}.
\newblock \bibinfo{journal}{\emph{Commun. {ACM}}} \bibinfo{volume}{55},
  \bibinfo{number}{3} (\bibinfo{year}{2012}), \bibinfo{pages}{40--44}.
\newblock


\bibitem[\protect\citeauthoryear{group}{group}{2017}]%
        {libfuzz}
\bibfield{author}{\bibinfo{person}{Libfuzzer group}.}
  \bibinfo{year}{2017}\natexlab{}.
\newblock \bibinfo{title}{LibFuzzer}.
\newblock \bibinfo{howpublished}{\url{https://llvm.org/docs/LibFuzzer.html}}.
\newblock


\bibitem[\protect\citeauthoryear{Gulzar, Zhu, and Han}{Gulzar
  et~al\mbox{.}}{2019}]%
        {DBLP:conf/icse/GulzarZH19}
\bibfield{author}{\bibinfo{person}{Muhammad~Ali Gulzar},
  \bibinfo{person}{Yongkang Zhu}, {and} \bibinfo{person}{Xiaofeng Han}.}
  \bibinfo{year}{2019}\natexlab{}.
\newblock \showarticletitle{Perception and practices of differential testing}.
  In \bibinfo{booktitle}{\emph{Proceedings of the 41st International Conference
  on Software Engineering: Software Engineering in Practice, {ICSE} {(SEIP)}
  2019, Montreal, QC, Canada, May 25-31, 2019}}. \bibinfo{publisher}{{IEEE} /
  {ACM}}, \bibinfo{pages}{71--80}.
\newblock


\bibitem[\protect\citeauthoryear{Hemmati}{Hemmati}{2015}]%
        {DBLP:conf/qrs/Hemmati15}
\bibfield{author}{\bibinfo{person}{Hadi Hemmati}.}
  \bibinfo{year}{2015}\natexlab{}.
\newblock \showarticletitle{How Effective Are Code Coverage Criteria?}. In
  \bibinfo{booktitle}{\emph{2015 {IEEE} International Conference on Software
  Quality, Reliability and Security, {QRS} 2015, Vancouver, BC, Canada, August
  3-5, 2015}}. \bibinfo{publisher}{{IEEE}}, \bibinfo{pages}{151--156}.
\newblock


\bibitem[\protect\citeauthoryear{Lemieux, Padhye, Sen, and Song}{Lemieux
  et~al\mbox{.}}{2018}]%
        {perffuzz}
\bibfield{author}{\bibinfo{person}{Caroline Lemieux}, \bibinfo{person}{Rohan
  Padhye}, \bibinfo{person}{Koushik Sen}, {and} \bibinfo{person}{Dawn Song}.}
  \bibinfo{year}{2018}\natexlab{}.
\newblock \showarticletitle{PerfFuzz: automatically generating pathological
  inputs}. In \bibinfo{booktitle}{\emph{Proceedings of the 27th {ACM} {SIGSOFT}
  International Symposium on Software Testing and Analysis, {ISSTA} 2018,
  Amsterdam, The Netherlands, July 16-21, 2018}}. \bibinfo{publisher}{{ACM}},
  \bibinfo{pages}{254--265}.
\newblock


\bibitem[\protect\citeauthoryear{Lou, Hao, and Zhang}{Lou
  et~al\mbox{.}}{2015}]%
        {DBLP:conf/issre/LouH015}
\bibfield{author}{\bibinfo{person}{Yiling Lou}, \bibinfo{person}{Dan Hao},
  {and} \bibinfo{person}{Lu Zhang}.} \bibinfo{year}{2015}\natexlab{}.
\newblock \showarticletitle{Mutation-based test-case prioritization in software
  evolution}. In \bibinfo{booktitle}{\emph{26th {IEEE} International Symposium
  on Software Reliability Engineering, {ISSRE} 2015, Gaithersbury, MD, USA,
  November 2-5, 2015}}. \bibinfo{publisher}{{IEEE} Computer Society},
  \bibinfo{pages}{46--57}.
\newblock


\bibitem[\protect\citeauthoryear{Ma, Cheng, Zhang, and Xuan}{Ma
  et~al\mbox{.}}{2020}]%
        {DBLP:journals/jss/MaCZX20}
\bibfield{author}{\bibinfo{person}{Ping Ma}, \bibinfo{person}{Hangyuan Cheng},
  \bibinfo{person}{Jingxuan Zhang}, {and} \bibinfo{person}{Jifeng Xuan}.}
  \bibinfo{year}{2020}\natexlab{}.
\newblock \showarticletitle{Can this fault be detected: {A} study on fault
  detection via automated test generation}.
\newblock \bibinfo{journal}{\emph{J. Syst. Softw.}}  \bibinfo{volume}{170}
  (\bibinfo{year}{2020}), \bibinfo{pages}{110769}.
\newblock


\bibitem[\protect\citeauthoryear{Ma, Yoo, and Kim}{Ma et~al\mbox{.}}{2021}]%
        {DBLP:conf/sigsoft/MaYK21}
\bibfield{author}{\bibinfo{person}{Yu-Seung Ma}, \bibinfo{person}{Shin Yoo},
  {and} \bibinfo{person}{Taeho Kim}.} \bibinfo{year}{2021}\natexlab{}.
\newblock \showarticletitle{Selecting test inputs for DNNs using differential
  testing with subspecialized model instances}. In
  \bibinfo{booktitle}{\emph{Proceedings of the 29th ACM Joint Meeting on
  European Software Engineering Conference and Symposium on the Foundations of
  Software Engineering}}. \bibinfo{pages}{1467--1470}.
\newblock


\bibitem[\protect\citeauthoryear{Man{\`{e}}s, Han, Han, Cha, Egele, Schwartz,
  and Woo}{Man{\`{e}}s et~al\mbox{.}}{2021}]%
        {fuzzSurvey}
\bibfield{author}{\bibinfo{person}{Valentin J.~M. Man{\`{e}}s},
  \bibinfo{person}{HyungSeok Han}, \bibinfo{person}{Choongwoo Han},
  \bibinfo{person}{Sang~Kil Cha}, \bibinfo{person}{Manuel Egele},
  \bibinfo{person}{Edward~J. Schwartz}, {and} \bibinfo{person}{Maverick Woo}.}
  \bibinfo{year}{2021}\natexlab{}.
\newblock \showarticletitle{The Art, Science, and Engineering of Fuzzing: {A}
  Survey}.
\newblock \bibinfo{journal}{\emph{{IEEE} Trans. Software Eng.}}
  \bibinfo{volume}{47}, \bibinfo{number}{11} (\bibinfo{year}{2021}),
  \bibinfo{pages}{2312--2331}.
\newblock


\bibitem[\protect\citeauthoryear{McKeeman}{McKeeman}{1998}]%
        {98difftest}
\bibfield{author}{\bibinfo{person}{William~M. McKeeman}.}
  \bibinfo{year}{1998}\natexlab{}.
\newblock \showarticletitle{Differential Testing for Software}.
\newblock \bibinfo{journal}{\emph{Digit. Tech. J.}} \bibinfo{volume}{10},
  \bibinfo{number}{1} (\bibinfo{year}{1998}), \bibinfo{pages}{100--107}.
\newblock


\bibitem[\protect\citeauthoryear{Miller, Fredriksen, and So}{Miller
  et~al\mbox{.}}{1990}]%
        {DBLP:journals/cacm/MillerFS90}
\bibfield{author}{\bibinfo{person}{Barton~P. Miller}, \bibinfo{person}{Lars
  Fredriksen}, {and} \bibinfo{person}{Bryan So}.}
  \bibinfo{year}{1990}\natexlab{}.
\newblock \showarticletitle{An Empirical Study of the Reliability of {UNIX}
  Utilities}.
\newblock \bibinfo{journal}{\emph{Commun. {ACM}}} \bibinfo{volume}{33},
  \bibinfo{number}{12} (\bibinfo{year}{1990}), \bibinfo{pages}{32--44}.
\newblock


\bibitem[\protect\citeauthoryear{Padhye, Lemieux, and Sen}{Padhye
  et~al\mbox{.}}{2019a}]%
        {jqf}
\bibfield{author}{\bibinfo{person}{Rohan Padhye}, \bibinfo{person}{Caroline
  Lemieux}, {and} \bibinfo{person}{Koushik Sen}.}
  \bibinfo{year}{2019}\natexlab{a}.
\newblock \showarticletitle{{JQF:} coverage-guided property-based testing in
  Java}. In \bibinfo{booktitle}{\emph{Proceedings of the 28th {ACM} {SIGSOFT}
  International Symposium on Software Testing and Analysis, {ISSTA} 2019,
  Beijing, China, July 15-19, 2019}}. \bibinfo{publisher}{{ACM}},
  \bibinfo{pages}{398--401}.
\newblock


\bibitem[\protect\citeauthoryear{Padhye, Lemieux, Sen, Papadakis, and
  Traon}{Padhye et~al\mbox{.}}{2019b}]%
        {zest}
\bibfield{author}{\bibinfo{person}{Rohan Padhye}, \bibinfo{person}{Caroline
  Lemieux}, \bibinfo{person}{Koushik Sen}, \bibinfo{person}{Mike Papadakis},
  {and} \bibinfo{person}{Yves~Le Traon}.} \bibinfo{year}{2019}\natexlab{b}.
\newblock \showarticletitle{Semantic fuzzing with zest}. In
  \bibinfo{booktitle}{\emph{Proceedings of the 28th {ACM} {SIGSOFT}
  International Symposium on Software Testing and Analysis, {ISSTA} 2019,
  Beijing, China, July 15-19, 2019}}. \bibinfo{publisher}{{ACM}},
  \bibinfo{pages}{329--340}.
\newblock


\bibitem[\protect\citeauthoryear{Papadakis, Kintis, Zhang, Jia, Traon, and
  Harman}{Papadakis et~al\mbox{.}}{2019}]%
        {mutestsurvey}
\bibfield{author}{\bibinfo{person}{Mike Papadakis}, \bibinfo{person}{Marinos
  Kintis}, \bibinfo{person}{Jie Zhang}, \bibinfo{person}{Yue Jia},
  \bibinfo{person}{Yves~Le Traon}, {and} \bibinfo{person}{Mark Harman}.}
  \bibinfo{year}{2019}\natexlab{}.
\newblock \showarticletitle{Chapter Six - Mutation Testing Advances: An
  Analysis and Survey}.
\newblock \bibinfo{journal}{\emph{Adv. Comput.}}  \bibinfo{volume}{112}
  (\bibinfo{year}{2019}), \bibinfo{pages}{275--378}.
\newblock


\bibitem[\protect\citeauthoryear{Papadakis and Traon}{Papadakis and
  Traon}{2015}]%
        {DBLP:journals/stvr/PapadakisT15}
\bibfield{author}{\bibinfo{person}{Mike Papadakis} {and}
  \bibinfo{person}{Yves~Le Traon}.} \bibinfo{year}{2015}\natexlab{}.
\newblock \showarticletitle{Metallaxis-FL: mutation-based fault localization}.
\newblock \bibinfo{journal}{\emph{Softw. Test. Verification Reliab.}}
  \bibinfo{volume}{25}, \bibinfo{number}{5-7} (\bibinfo{year}{2015}),
  \bibinfo{pages}{605--628}.
\newblock


\bibitem[\protect\citeauthoryear{Park, An, Youn, Kim, and Ryu}{Park
  et~al\mbox{.}}{2021}]%
        {DBLP:conf/icse/ParkAYKR21}
\bibfield{author}{\bibinfo{person}{Jihyeok Park}, \bibinfo{person}{Seungmin
  An}, \bibinfo{person}{Dongjun Youn}, \bibinfo{person}{Gyeongwon Kim}, {and}
  \bibinfo{person}{Sukyoung Ryu}.} \bibinfo{year}{2021}\natexlab{}.
\newblock \showarticletitle{{JEST:} {N+1} -version Differential Testing of Both
  JavaScript Engines and Specification}. In \bibinfo{booktitle}{\emph{43rd
  {IEEE/ACM} International Conference on Software Engineering, {ICSE} 2021,
  Madrid, Spain, 22-30 May 2021}}. \bibinfo{publisher}{{IEEE}},
  \bibinfo{pages}{13--24}.
\newblock


\bibitem[\protect\citeauthoryear{Shi, Gyori, Gligoric, Zaytsev, and
  Marinov}{Shi et~al\mbox{.}}{2014}]%
        {DBLP:conf/sigsoft/ShiGGZM14}
\bibfield{author}{\bibinfo{person}{August Shi}, \bibinfo{person}{Alex Gyori},
  \bibinfo{person}{Milos Gligoric}, \bibinfo{person}{Andrey Zaytsev}, {and}
  \bibinfo{person}{Darko Marinov}.} \bibinfo{year}{2014}\natexlab{}.
\newblock \showarticletitle{Balancing trade-offs in test-suite reduction}. In
  \bibinfo{booktitle}{\emph{Proceedings of the 22nd {ACM} {SIGSOFT}
  International Symposium on Foundations of Software Engineering, (FSE-22),
  Hong Kong, China, November 16 - 22, 2014}}. \bibinfo{publisher}{{ACM}},
  \bibinfo{pages}{246--256}.
\newblock


\bibitem[\protect\citeauthoryear{Sun, Wang, Wu, Duan, Sun, and Chen}{Sun
  et~al\mbox{.}}{2019}]%
        {sun2019maf}
\bibfield{author}{\bibinfo{person}{Weisong Sun}, \bibinfo{person}{Xingya Wang},
  \bibinfo{person}{Haoran Wu}, \bibinfo{person}{Ding Duan},
  \bibinfo{person}{Zesong Sun}, {and} \bibinfo{person}{Zhenyu Chen}.}
  \bibinfo{year}{2019}\natexlab{}.
\newblock \showarticletitle{MAF: method-anchored test fragmentation for test
  code plagiarism detection}. In \bibinfo{booktitle}{\emph{2019 IEEE/ACM 41st
  International Conference on Software Engineering: Software Engineering
  Education and Training (ICSE-SEET)}}. IEEE, \bibinfo{pages}{110--120}.
\newblock


\bibitem[\protect\citeauthoryear{Wen, Wang, Li, Qin, Liu, Xu, Chen, Xie, Pu,
  and Liu}{Wen et~al\mbox{.}}{2020}]%
        {memlock}
\bibfield{author}{\bibinfo{person}{Cheng Wen}, \bibinfo{person}{Haijun Wang},
  \bibinfo{person}{Yuekang Li}, \bibinfo{person}{Shengchao Qin},
  \bibinfo{person}{Yang Liu}, \bibinfo{person}{Zhiwu Xu},
  \bibinfo{person}{Hongxu Chen}, \bibinfo{person}{Xiaofei Xie},
  \bibinfo{person}{Geguang Pu}, {and} \bibinfo{person}{Ting Liu}.}
  \bibinfo{year}{2020}\natexlab{}.
\newblock \showarticletitle{MemLock: memory usage guided fuzzing}. In
  \bibinfo{booktitle}{\emph{{ICSE} '20: 42nd International Conference on
  Software Engineering, Seoul, South Korea, 27 June - 19 July, 2020}}.
  \bibinfo{publisher}{{ACM}}, \bibinfo{pages}{765--777}.
\newblock


\bibitem[\protect\citeauthoryear{Ye, Tang, Tan, Huang, Fang, Sun, Bian, Wang,
  and Wang}{Ye et~al\mbox{.}}{2021}]%
        {DBLP:conf/pldi/YeTTHFSBW021}
\bibfield{author}{\bibinfo{person}{Guixin Ye}, \bibinfo{person}{Zhanyong Tang},
  \bibinfo{person}{Shin~Hwei Tan}, \bibinfo{person}{Songfang Huang},
  \bibinfo{person}{Dingyi Fang}, \bibinfo{person}{Xiaoyang Sun},
  \bibinfo{person}{Lizhong Bian}, \bibinfo{person}{Haibo Wang}, {and}
  \bibinfo{person}{Zheng Wang}.} \bibinfo{year}{2021}\natexlab{}.
\newblock \showarticletitle{Automated conformance testing for JavaScript
  engines via deep compiler fuzzing}. In \bibinfo{booktitle}{\emph{Proceedings
  of the 42nd ACM SIGPLAN International Conference on Programming Language
  Design and Implementation}}. \bibinfo{pages}{435--450}.
\newblock


\bibitem[\protect\citeauthoryear{Yun, Lee, Xu, Jang, and Kim}{Yun
  et~al\mbox{.}}{2018}]%
        {qsym}
\bibfield{author}{\bibinfo{person}{Insu Yun}, \bibinfo{person}{Sangho Lee},
  \bibinfo{person}{Meng Xu}, \bibinfo{person}{Yeongjin Jang}, {and}
  \bibinfo{person}{Taesoo Kim}.} \bibinfo{year}{2018}\natexlab{}.
\newblock \showarticletitle{{QSYM} : {A} Practical Concolic Execution Engine
  Tailored for Hybrid Fuzzing}. In \bibinfo{booktitle}{\emph{27th {USENIX}
  Security Symposium, {USENIX} Security 2018, Baltimore, MD, USA, August 15-17,
  2018}}. \bibinfo{publisher}{{USENIX} Association}, \bibinfo{pages}{745--761}.
\newblock


\bibitem[\protect\citeauthoryear{Zalewski}{Zalewski}{2017}]%
        {afl}
\bibfield{author}{\bibinfo{person}{Michal Zalewski}.}
  \bibinfo{year}{2017}\natexlab{}.
\newblock \bibinfo{title}{American Fuzzy Lop 2.5.2b}.
\newblock \bibinfo{howpublished}{\url{https://lcamtuf.coredump.cx/afl/}}.
\newblock


\end{thebibliography}

\end{document}